\documentclass{JHEP3}
\usepackage{epsfig}
\usepackage{amsbsy}
\usepackage{varioref}
\usepackage{pifont}
\usepackage{axodraw}

\newcommand{\hs}{\hspace*{0.5cm}}
\newcommand{\vs}{\vspace*{0.5cm}}
\newcommand{\be}{\begin{equation}}
\newcommand{\ee}{\end{equation}}
\newcommand{\bea}{\begin{eqnarray}}
\newcommand{\eea}{\end{eqnarray}}
\newcommand{\bary}{\begin{array}}
\newcommand{\eary}{\end{array}}
\newcommand{\bit}{\begin{itemize}}
\newcommand{\eit}{\end{itemize}}
\newcommand{\ben}{\begin{enumerate}}
\newcommand{\een}{\end{enumerate}}
\newcommand{\crn}{\nonumber \\}
\newcommand{\nn}{\nonumber}

\newcommand{\al}{\alpha}
\newcommand{\la}{\lambda}
\newcommand{\bet}{\beta}
\newcommand{\ga}{\gamma}

\newcommand{\om}{\omega}
\newcommand{\pa}{\partial}
\newcommand{\fr}{\frac}

\newcommand{\bc}{\begin{center}}
\newcommand{\ec}{\end{center}}

\newcommand{\de}{\delta}
\newcommand{\De}{\Delta}
\newcommand{\ep}{\epsilon}

\newcommand{\si}{\sigma}

\def\sla#1{\ifmmode%
\setbox0=\hbox{$#1$}%
\setbox1=\hbox to\wd0{\hss$/$\hss}\else%
\setbox0=\hbox{#1}%
\setbox1=\hbox to\wd0{\hss/\hss}\fi%
#1\hskip-\wd0\box1 }
\title{Sfermion masses  in the supersymmetric economical
3-3-1 model}
\author{ P. V.  Dong, Tr. T. Huong,  N.  T.
Thuy and  H. N. Long \\   Institute of  Physics, VAST, P. O. Box
429, Bo Ho, Hanoi
10000, Vietnam\\
 E-mail:
 \email{pvdong@iop.vast.ac.vn},\email{tthuong@iop.vast.ac.vn},
 \email{ntthuyy7@yahoo.com.vn},
\email{hnlong@iop.vast.ac.vn}}

 \abstract{Sfermion  masses and eigenstates in
 the supersymmetric  economical
3-3-1 model are studied. By lepton number conservation, the exotic
squarks and superpartners of ordinary quarks are decoupled.  Due to
the fact that in the 3-3-1 models, one generation of quarks behaves
differently from other two, by $R$-parity conservation, the mass
mixing matrix of the squarks in this model are smaller than that in
the Minimal Supersymmetric Standard Model (MSSM). Assuming
substantial mixing in pairs of highest flavours, we are able to get
mass spectrum and eigenstates of all the sfermions. In the effective
approximation, the slepton  mass splittings in the first two
generations, are consistent with those in the MSSM, namely: $
m^2_{\tilde{l}_L} - m^2_{\tilde{\nu}_{l L}}  = m_W^2 \cos 2\ga$
$(l=e,\ \mu)$. In addition, within the above effective limit, there
exists degeneracy among sneutrinos in each multiplet:
$m^2_{\tilde{\nu}_{l L}} = m^2_{\tilde{\nu}_{l R}}$. In
contradiction to the MSSM, the squark mass splittings are different
for each generation and not to be $ m_W^2 \cos 2\ga$. }


\begin{document}

\maketitle

\section{\label{intro}Introduction}

The Standard Model (SM) of high energy physics provides a remarkable
successful  description of presently known phenomena. In spite of
these successes, it fails to explain several fundamental issues like
generation number puzzle, neutrino masses and oscillations, the
origin of charge quantization, CP violation, etc.

One of the simplest solutions to these problems is to enhance the
SM symmetry  $\mathrm{SU}(3)_{C} \otimes \mathrm{SU}(2)_{L}
\otimes \mathrm{U}(1)_{Y}$ to $\mathrm{SU}(3)_{C} \otimes
\mathrm{SU}(3)_{L} \otimes \mathrm{U}(1)_{X}$ (called 3-3-1 for
short)~\cite{ppf,flt,331rh}  gauge group.
 One of the main motivations to study this kind of models
is an explanation in part of the generation number puzzle. In the
3-3-1 models, each generation is not anomaly free; and the model
becomes anomaly free if one of quark families behaves differently
from other two. Consequently, the number of generations is multiple
of the color number. Combining with the QCD asymptotic freedom, the
generation number has to be three. For the neutrino masses and
oscillations, the electric charge quantization and CP violation
issues in the 3-3-1 models, the interested readers can find in Refs.
\cite{neu331}, \cite{chargequan} and \cite{CP331}, respectively.

In one of the 3-3-1 models, the right-handed neutrinos are in bottom
of the lepton triplets \cite{331rh} and three Higgs triplets are
required. It is worth noting that, there are two Higgs triplets with
{\it neutral components in the top and bottom}. In the earlier
version, these triplets can have vacuum expectation value (VEV)
either on the top or in the bottom, but not in both. Assuming that
all neutral components in the triplet can have VEVs, we are able to
reduce number of triplets in the model to be two
~\cite{ponce,haihiggs}. Such a scalar sector is minimal, therefore
it has been called the economical 3-3-1 model~\cite{higgseconom}. In
a series of papers, we have developed and proved that this
non-supersymmetric version is consistent, realistic and very rich in
physics \cite{haihiggs,higgseconom, dlhh,dls1}.

In the other hands, due to the ``no-go" theorem of Coleman-Mandula
\cite{nogo}, the internal $G$ and external $P$ spacetime symmetries
can only be {\it trivially} unified. In addition, the mere fact that
the ratio $M_P/M_W$ is so huge is already a powerful clue to the
character of physics beyond the SM, because of the infamous
hierarchy problem. In the framework of new symmetry called a
supersymmetry \cite{susy,martin}, the above mentioned problems can
be solved. One of the intriguing features of supersymmetric theories
is that the Higgs spectrum (unfortunately, the only part of the SM
is still not discovered) is quite constrained.

It is known that the economical (non-supersymmetric) 3-3-1 model
does not furnish any candidate for self-interaction dark
matter~\cite{SIDM} with the condition given by Spergel and
Steinhardt~\cite{ss}. With a larger content of the scalar sector,
the supersymmetric version is expected to have a candidate for the
self-interaction dark matter. The supersymmetric version of the
3-3-1 model with right-handed neutrinos~\cite{331rh} has already
been constructed in Refs.~\cite{s331r}. An supersymmetric version of
the economical 3-3-1 model has been constructed in Ref.
\cite{susyeco}. Some interesting features such as Higgs bosons with
masses equal to that of the gauge bosons -- the $W$ and the
bileptons $X$ and $Y$, have been pointed out in Ref. \cite{higph}.

In a supersymmetric extension of the (beyond) SM, each of the known
fundamental particles must be in either a chiral or gauge
supermultiplet and have a superpartner with spin differing by 1/2
unit. All of the matter fermions (the known quarks and leptons) have
spin-0 partners called sfermions. Hence in supersymmetric models,
besides scalar Higgs bosons, there are scalar sfermions. In Ref.
\cite{susyeco}, the Higgs sector was a subject of our interest; in
this paper, we will focus an attention to the sfermions -- sleptons
and squarks.

This article is organized as follows. In Sec.~\ref{model} we present
a fermion and scalar content in the supersymmetric economical 3-3-1
model. The necessary parts of Lagrangian is also given. The $F$ and
$D$ terms of scalar potential for sfermions are calculated in
Sec.~\ref{scalarpot}. Masses and eigenstates for sleptons and
squarks are given in Sec.~\ref{slepm} and  \ref {squarkm},
respectively. Section \ref{rparity} is devoted for the case of
$R$-parity conservation and sfermion  mass splittings. Finally, we
summarize our results and make conclusions in the last section -
Sec.~\ref{concl}.

\section{A review of the model}
\label{model} In this section we first recapitulate the basic
elements of the supersymmetric economical 3-3-1 model
\cite{susyeco}. $R-parity$ and some constraints on the couplings are
also presented.

\subsection{\label{parcontent}Particle content }

The superfield content in this paper is defined in a standard way as
follows \be \widehat{F}= (\widetilde{F}, F),\hs \widehat{S} = (S,
\widetilde{S}),\hs \widehat{V}= (\lambda,V), \ee where the
components $F$, $S$ and $V$ stand for the fermion, scalar and vector
fields while their superpartners are denoted as $\widetilde{F}$,
$\widetilde{S}$ and $\lambda$, respectively \cite{susy,s331r}.

The superfield content in the considering model with an anomaly-free
fermionic content transforms under the 3-3-1 gauge group as
\begin{equation}
\widehat{L}_{a L}=\left(\widehat{\nu}_{a}, \widehat{l}_{a},
\widehat{\nu}^c_{a}\right)^T_{L} \sim (1,3,-1/3),\hs
  \widehat {l}^{c}_{a L} \sim (1,1,1),\label{l2}
\end{equation}  \be \widehat Q_{1L}=
\left(\widehat { u}_1,\
                        \widehat {d}_1,\
                        \widehat {u}^\prime
 \right)^T_L \sim (3,3,1/3),\label{quarks3}\ee
\be \widehat {u}^{c}_{1L},\ \widehat { u}^{ \prime c}_{L} \sim
(3^*,1,-2/3),\hs \widehat {d}^{c}_{1L} \sim (3^*,1,1/3 ), \label{l5}
\ee
\begin{equation}
\begin{array}{ccc}
 \widehat{Q}_{\alpha L} = \left(\widehat{d}_{\alpha},
 - \widehat{u}_{\alpha},
 \widehat{d^\prime}_{\alpha}\right)^T_{L}
 \sim (3,3^*,0), \hs \al=2,3, \label{l3}
\end{array}
\end{equation}
\begin{equation}
\widehat{u}^{c}_{\alpha L} \sim \left(3^*,1,-2/3 \right),\hs
\widehat{d}^{c}_{\alpha L},\ \widehat{d}^{\prime c}_{\alpha L} \sim
\left(3^*,1,1/3 \right), \label{l4}
\end{equation}
where the values in the parentheses denote quantum numbers based on
$\left(\mbox{SU}(3)_C\right.$, $\mbox{SU}(3)_L$,
$\left.\mbox{U}(1)_X\right)$ symmetry.
$\widehat{\nu}^c_L=(\widehat{\nu}_R)^c$ and $a=1,2,3$ is a
generation index. The primes superscript on usual quark types ($u'$
with the electric charge $q_{u'}=2/3$ and $d'$ with $q_{d'}=-1/3$)
indicate that those quarks are exotic ones.

The two superfields $\widehat{\chi}$ and $\widehat {\rho} $ are at
least introduced to span the scalar sector of the economical 3-3-1
model \cite{higgseconom}: \bea \widehat{\chi}&=& \left (
\widehat{\chi}^0_1, \widehat{\chi}^-, \widehat{\chi}^0_2
\right)^T\sim (1,3,-1/3), \label{l7}\\
\widehat{\rho}&=& \left (\widehat{\rho}^+_1, \widehat{\rho}^0,
\widehat{\rho}^+_2\right)^T \sim  (1,3,2/3). \label{l8} \eea To
cancel the chiral anomalies of Higgsino sector, the two extra
superfields $\widehat{\chi}^\prime$ and $\widehat {\rho}^\prime $
must be added as follows \bea \widehat{\chi}^\prime&=& \left
(\widehat{\chi}^{\prime 0}_1, \widehat{\chi}^{\prime
+},\widehat{\chi}^{\prime 0}_2 \right)^T\sim ( 1,3^*,1/3),
\label{l9}\\
\widehat{\rho}^\prime &=& \left (\widehat{\rho}^{\prime -}_1,
  \widehat{\rho}^{\prime 0},  \widehat{\rho}^{\prime -}_2
\right)^T\sim (1,3^*,-2/3). \label{l10} \eea

In this model, the $ \mathrm{SU}(3)_L \otimes \mathrm{U}(1)_X$
gauge group is broken via two steps:
 \be \mathrm{SU}(3)_L \otimes
\mathrm{U}(1)_X \stackrel{w,w'}{\longrightarrow}\ \mathrm{SU}(2)_L
\otimes \mathrm{U}(1)_Y\stackrel{v,v',u,u'}{\longrightarrow}
\mathrm{U}(1)_{Q},\label{stages}\ee where the VEVs are defined by
\bea
 \sqrt{2} \langle\chi\rangle^T &=& \left(u, 0, w\right), \hs \sqrt{2}
 \langle\chi^\prime\rangle^T = \left(u^\prime,  0,
 w^\prime\right),\\
\sqrt{2}  \langle\rho\rangle^T &=& \left( 0, v, 0 \right), \hs
\sqrt{2} \langle\rho^\prime\rangle^T = \left( 0, v^\prime,  0
\right).\eea The VEVs $w$ and $w^\prime$ are responsible for the
first step of the symmetry breaking while $u,\ u^\prime$ and $v,\
v^\prime$ are for the second one. Therefore, they have to satisfy
the constraints:
 \be
 u,\ u^\prime,\ v,\ v^\prime
\ll w,\ w^\prime. \label{contraint}\ee

The vector superfields $\widehat{V}_c$, $\widehat{V}$ and
$\widehat{V}^\prime$ containing the usual gauge bosons are,
respectively, associated with the $\mathrm{SU}(3)_C$,
$\mathrm{SU}(3)_L$ and $\mathrm{U}(1)_X $ group factors. The colour
and flavour vector superfields have expansions in the Gell-Mann
matrix bases $T^a=\lambda^a/2$ $(a=1,2,...,8)$ as follows\bea
\widehat{V}_c &=& \fr{1}{2}\lambda^a \widehat{V}_{ca},\hs
\widehat{\overline{V}}_c=-\fr{1}{2}\lambda^{a*} \widehat{V}_{ca};\hs
\widehat{V} = \fr{1}{2}\lambda^a \widehat{V}_{a},\hs
\widehat{\overline{V}}=-\fr{1}{2}\lambda^{a*} \widehat{V}_{a},\eea
where an overbar $^-$ indicates complex conjugation. For the vector
superfield associated with $\mathrm{U}(1)_X$, we normalize as
follows \be X \hat{V}'= (XT^9) \hat{B}, \hs
T^9\equiv\fr{1}{\sqrt{6}}\mathrm{diag}(1,1,1).\ee The gluons are
denoted by $g^a$ and their respective gluino partners by
$\lambda^a_{c}$, with $a=1, \ldots,8$. In the electroweak sector,
$V^a$ and $B$ stand for the $\mathrm{SU}(3)_{L}$ and
$\mathrm{U}(1)_{X}$ gauge bosons with their gaugino partners
$\lambda^a_{V}$ and $\lambda_{B}$, respectively.

With the superfields as given, the full Lagrangian is defined by
$\mathcal{L}_{susy}+\mathcal{L}_{soft}$, where the first term is
supersymmetric part, whereas the last term breaks explicitly the
supersymmetry \cite{susyeco}. The interested reader can find more
details on this Lagrangian in the above mentioned article. In the
following, only terms relevant to our calculations are displayed.

From the supersymmetric Lagrangian \cite{susyeco}, we can obtain the
following superpotential
 \begin{equation} W= \frac{W_{2}}{2}+ \frac{W_{3}}{3},
\end{equation}
where
\begin{eqnarray}
W_{2}&=& \mu_{0a}\hat{L}_{aL} \hat{ \chi}^{\prime}+ \mu_{ \chi}
\hat{ \chi} \hat{ \chi}^{\prime}+
 \mu_{ \rho} \hat{ \rho} \hat{ \rho}^{\prime},
 \label{w2}
\end{eqnarray}
and
\begin{eqnarray} W_{3}&=&
\ga_{ab} \hat{L}_{aL} \hat{ \rho}^{\prime} \hat{l}^{c}_{bL}+
\la_{a} \epsilon \hat{L}_{aL} \hat{\chi} \hat{\rho}+
\la^\prime_{ab} \epsilon \hat{L}_{aL} \hat{L}_{bL}
\hat{\rho} \nonumber \\
&&+ \kappa_{i} \hat{Q}_{1L} \hat{\chi}^{\prime} \hat{u}^{c}_{iL}+
\kappa^\prime \hat{Q}_{1L} \hat{\chi}^{\prime} \hat{u}^{\prime
c}_L+ \vartheta_{i}\hat{Q}_{1L} \hat{\rho}^{\prime}
\hat{d}^{c}_{iL} \nonumber \\
&&+ \vartheta^\prime_{ \alpha}\hat{Q}_{1L} \hat{\rho}^{\prime}
\hat{d}^{\prime c}_{\alpha L} + \pi_{ \alpha i} \hat{Q}_{\alpha
L}\hat{\rho}\hat{u}^{c}_{iL} +\pi_{\alpha}^{\prime}
\hat{Q}_{\alpha L}\hat{\rho}\hat{u}^{\prime c}_{L} \nonumber \\
&&+ \Pi_{\alpha i} \hat{Q}_{\alpha L} \hat{\chi} \hat{d}^{c}_{iL}
+ \Pi^\prime_{\alpha \beta} \hat{Q}_{\alpha L} \hat{\chi}
\hat{d}^{\prime c}_{\beta L}+ \epsilon
f_{\alpha\beta\gamma}\hat{Q}_{\alpha L} \hat{Q}_{\beta L}
\hat{Q}_{\gamma L} \crn &&+ \xi_{1i \beta j} \hat{d}^{c}_{iL}
\hat{d}^{\prime c}_{\beta L} \hat{u}^{c}_{j L}+ \xi_{2i \beta }
\hat{d}^{c}_{i L} \hat{d}^{\prime c}_{\beta L} \hat{u}^{\prime
c}_{L}+ \xi_{3ijk}
\hat{d}^{c}_{iL} \hat{d}^{c}_{jL} \hat{u}^{c}_{k L} \nonumber \\
&&+ \xi_{4ij} \hat{d}^{c}_{i L} \hat{d}^{c}_{jL} \hat{u}^{\prime
c}_{L}+ \xi_{5 \alpha \beta i} \hat{d}^{\prime c}_{\alpha L}
\hat{d}^{\prime c}_{\beta L} \hat{u}^{c}_{iL} + \xi_{6 \alpha
\beta} \hat{d}^{\prime c}_{\alpha L}\hat{d}^{\prime c}_{\beta L}
\hat{u}^{\prime c}_{L} \crn &&+ \xi_{a \alpha j}\hat{L}_{aL}
\hat{Q}_{\alpha L} \hat{d}^{c}_{jL}+ \xi^\prime_{a\alpha
\beta}\hat{L}_{aL} \hat{Q}_{\alpha L} \hat{d}^{\prime c}_{\beta
L}. \label{w3}
\end{eqnarray}
The coefficients $\mu_{0a}, \mu_{\rho}$ and $\mu_{\chi}$ have mass
dimension, while all  coefficients in $W_{3}$ are dimensionless
and $\la^\prime_{ab}= - \la^\prime_{ba}$.

It is worth noting that the first term of (\ref{w3}) is the Yukawa
coupling giving charged leptons  mass, while the third one is
responsible for neutrino mass. At the tree level,  their couplings
satisfy the following estimation \cite{dls1,marcos}: \be \ga_{ab}
 \gg \la^\prime_{ab}. \label{hsyln} \ee In the SM,
neutrinos are rigid  massless, hence  $\la^\prime_{ab}$ has to be
vanish. In other words, we can put $\la^\prime_{ab} = 0$ in the SM
limit.

From the soft supersymmetry-breaking terms \cite{susyeco}, the
Lagangian relevant to the sfermions is obtained by \bea
-\mathcal{L}_{SMT}&=& M^2_{ab}\widetilde{L}_{aL}^\dagger
\widetilde{L}_{bL}+ m^2_{ab}\widetilde{l}_{aL}^{c *}
\widetilde{l}_{bL}^c +m^2_{Q1L}\widetilde{Q}_{1L}^\dagger
\widetilde{Q}_{1 L}+m^2_{Q\alpha \bet L}\widetilde{Q}_{\alpha
L}^\dagger \widetilde{Q}_{\bet L}\crn &&+
m^2_{u_{ij}}\widetilde{u}_{iL}^{c
*} \widetilde{u}_{jL}^c + m^2_{d_{ij}}\widetilde{d}_{iL}^{c *}
\widetilde{d}_{jL}^c + m^2_{u^\prime }\widetilde{u^\prime}_{L}^{c
*} \widetilde{u^\prime}_{L}^c+ m^2_{d^\prime \al
\bet}\widetilde{d^\prime}_{\al L }^{c *} \widetilde{d^\prime}_{\bet
L}^c \crn && +\left\{M^{\prime 2}_a
\chi^\dagger\widetilde{L}_{aL}\right.
+\eta_{ab}\widetilde{L}_{aL}\rho ^\prime
\widetilde{l}_{Lb}^c+\upsilon_{a}\epsilon\widetilde{L}_{aL}\chi\rho+
\varepsilon_{ab}\epsilon \widetilde{L}_{aL}\widetilde{L}_{bL}\rho
+p_{i}\widetilde{Q}_{1L}\chi^\prime\widetilde{u}^c_{iL} \crn
&&+p\widetilde{Q}_{1L}\chi^\prime\widetilde{ u^\prime}^c_{L} + p_{
\alpha i}\widetilde{Q}_{\alpha
L}\rho\widetilde{u}^c_{iL}+r_{\al}\widetilde{Q}_{\alpha
L}\rho\widetilde{u^\prime}^c_{L}+
h_{i}\widetilde{Q}_{1L}\rho^\prime\widetilde{d}^c_{iL} \crn &&+
h^\prime_{i}\widetilde{Q}_{1L}\rho^\prime\widetilde{d^\prime}^c_{i
L} +h_{\al i}\widetilde{Q}_{\alpha L}\chi\widetilde{d}^c_{iL}+
h^\prime_{\alpha\beta}\widetilde{Q}_{\alpha
L}\chi\widetilde{d^\prime}^c_{\beta L}\crn && +
p_{5\alpha\beta\gamma}\widetilde{Q}_{\alpha L}
\widetilde{Q}_{{\beta} L}\widetilde{Q}_{{\gamma} L} +\kappa_{i\beta
j}\widetilde{d}_{iL}^c\widetilde{d^\prime}_{\beta
L}^c\widetilde{u}_{jL}^c+\vartheta_{i\beta
}\widetilde{d}_{iL}^c\widetilde{d^\prime}_{\beta
L}^c\widetilde{u^\prime}_{L}^c \nonumber \\ && +\pi_{i j
k}\widetilde{d}_{iL}^c\widetilde{d}_{j L}^c\widetilde{u}_{kL}^c
+\kappa_{4i k}\widetilde{d}_{iL}^c\widetilde{d}_{j
L}^c\widetilde{u^\prime}_{L}^c +\kappa_{5\alpha \beta
i}\widetilde{d^\prime}_{\alpha L}^c\widetilde{d^\prime}_{\beta
L}^c\widetilde{u}_{i L}^c \crn &&\left. + \kappa_{6\alpha \beta
}\widetilde{d^\prime}_{\alpha L}^c\widetilde{d^\prime}_{\beta
L}^c\widetilde{u^\prime}_{ L}^c + \om_{a\alpha j }\widetilde{L}_{a
L}\widetilde{Q}_{\alpha L}\widetilde{d}_{j L}^c+\om^\prime_{a\alpha
\beta }\widetilde{L}_{a L}\widetilde{Q}_{\alpha
L}\widetilde{d^\prime}_{\beta L}^c+ H.c.\right\},  \label{mme}\eea
where  $\varepsilon_{ab}= - \varepsilon_{ba}$. This Lagrangian is
also responsible for sfermion masses.

\subsection{$R$-parity} For the further analysis, it is convenience
to introduce $R$-parity in the model.  Following Ref. \cite{marcos},
$R$-parity can be expressed as follows
\begin{equation}
R-parity=(-1)^{2S}(-1)^{3({\cal B}+{\cal L})}
\label{rfor}\end{equation} where invariant charges ${\cal L}$ and
${\cal B}$ (for details, see  Ref. \cite{changlong}) are given by
\begin{equation}
\begin{array}{|c|c|c|c|c|}
\hline
  Triplet & L & Q_{1} & \chi  & \rho \\
  \hline
  {\cal B} \,\  charge & 0 & \frac{1}{3} & 0  & 0 \\ \hline
  {\cal L} \,\  charge & \frac{1}{3} & - \frac{2}{3} & \frac{4}{3}
  & - \frac{2}{3} \\ \hline
\end{array}
\end{equation}
\begin{equation}
\begin{array}{|c|c|c|c|}
\hline
  Anti-Triplet & Q_{\alpha} & \chi^{\prime}  & \rho^{\prime} \\
  \hline
  {\cal B} \,\  charge & \frac{1}{3} & 0 &  0 \\ \hline
  {\cal L} \,\  charge &  \frac{2}{3} & - \frac{4}{3}
  & \frac{2}{3} \\ \hline
\end{array}
\end{equation}
\begin{equation}
\begin{array}{|c|c|c|c|c|c|}
\hline
  Singlet & l^{c} & u^{c} & d^{c} & u^{\prime c} & d^{\prime c} \\
  \hline
  {\cal B} \,\ charge & 0 &- \frac{1}{3} &
  - \frac{1}{3} & - \frac{1}{3} & - \frac{1}{3} \\ \hline
  {\cal L} \,\ charge & - 1 & 0 &  0 & 2 & -2 \\ \hline
\end{array}
\end{equation}

Combining (\ref{rfor}) and the above tables, it is easy to conclude
that the fields $ \chi $, $ \chi^{\prime}$, $\rho $,
$\rho^{\prime}$, $L$, $Q_{\alpha}$, $Q_{3}$, $l$, $u$, $u^{\prime}$,
$d$ and $d^{\prime}$ have $R$-charge equal to one, while their
superpartners have opposite $R$-charge, as in the Minimal
Supersymmetric Standard Model (MSSM).

Under $R$-parity transformation, the Higgs and matter superfields
change, respectively~\cite{marcos}:
 \bea
\hat{H}_{1,2}(x,\theta,\bar{\theta}) &\stackrel{{\bf
R}_{d}}{\longmapsto}& \hat{H}_{1,2}(x,-\theta,-\bar{\theta} ),
\crn
 \hat{\Phi}(x,\theta,\bar{\theta}) &\stackrel{{\bf
R}_{d}}{\longmapsto}& - \hat{\Phi}(x,-\theta,-\bar{\theta} ),
 \ \Phi=Q,u^c,d^c,L,l^c,
\label{Rpa1c} \eea

Let us separate $W$ and $\mathcal{L}_{SMT}$ into the $R$-parity
conserving ($R$) and violating ($R\!\!\!\!/$) part. Thus \be W =
W_{R} + W_{R\!\!\!\!/}, \label{Rw} \ee where \bea W_{R} & = & \fr 1
2 \left(\mu_{ \chi} \hat{ \chi} \hat{ \chi}^{\prime}+
 \mu_{ \rho} \hat{ \rho} \hat{ \rho}^{\prime}\right)\crn
 & & + \fr
1 3 \left(\ga_{ab} \hat{L}_{aL} \hat{ \rho}^{\prime}
\hat{l}^{c}_{bL}+ \la^\prime_{ab} \epsilon \hat{L}_{aL}
\hat{L}_{bL}
\hat{\rho} \right.\nonumber \\
&&+ \kappa^\prime \hat{Q}_{1L} \hat{\chi}^{\prime} \hat{u}^{\prime
c}_L+ \vartheta_{i}\hat{Q}_{1L} \hat{\rho}^{\prime} \hat{d}^{c}_{iL}
+ \pi_{ \alpha i} \hat{Q}_{\alpha L}\hat{\rho}\hat{u}^{c}_{iL}+
\Pi_{\alpha i} \hat{Q}_{\alpha L} \hat{\chi}
\hat{d}^{c}_{iL} \nonumber \\
&&+ \kappa_{i} \hat{Q}_{1L} \hat{\chi}^{\prime} \hat{u}^{c}_{iL}+
\vartheta^\prime_{ \alpha}\hat{Q}_{1L} \hat{\rho}^{\prime}
\hat{d}^{\prime c}_{\alpha L} +\pi_{\alpha}^{\prime} \hat{Q}_{\alpha
L}\hat{\rho}\hat{u}^{\prime c}_{L} + \left. \Pi^\prime_{\alpha
\beta} \hat{Q}_{\alpha L} \hat{\chi} \hat{d}^{\prime c}_{\beta
L}\right), \label{Rw1} \eea and \bea W_{R\!\!\!\!/} & = & \fr 1 2
\mu_{0a}\hat{L}_{aL} \hat{ \chi}^{\prime} + \fr 1 3 \left( \la_{a}
\epsilon \hat{L}_{aL} \hat{\chi} \hat{\rho}+ \epsilon
f_{\alpha\beta\gamma}\hat{Q}_{\alpha L} \hat{Q}_{\beta L}
\hat{Q}_{\gamma L}\right. \crn &&+ \xi_{1i \beta j} \hat{d}^{c}_{iL}
\hat{d}^{\prime c}_{\beta L} \hat{u}^{c}_{j L}+ \xi_{2i \beta }
\hat{d}^{c}_{i L} \hat{d}^{\prime c}_{\beta L} \hat{u}^{\prime
c}_{L}+ \xi_{3ijk}
\hat{d}^{c}_{iL} \hat{d}^{c}_{jL} \hat{u}^{c}_{k L} \nonumber \\
&&+ \xi_{4ij} \hat{d}^{c}_{i L} \hat{d}^{c}_{jL} \hat{u}^{\prime
c}_{L}+ \xi_{5 \alpha \beta i} \hat{d}^{\prime c}_{\alpha L}
\hat{d}^{\prime c}_{\beta L} \hat{u}^{c}_{iL} + \xi_{6 \alpha \beta}
\hat{d}^{\prime c}_{\alpha L}\hat{d}^{\prime c}_{\beta L}
\hat{u}^{\prime c}_{L} \crn &&+\left. \xi_{a \alpha j}\hat{L}_{aL}
\hat{Q}_{\alpha L} \hat{d}^{c}_{jL}+ \xi^\prime_{a\alpha
\beta}\hat{L}_{aL} \hat{Q}_{\alpha L} \hat{d}^{\prime c}_{\beta
L}\right). \label{wrv}\eea By (\ref{Rpa1c}), the  $R\!\!\!\!/$ part
contains odd number of {\it matter} superfields. For the soft terms,
we have also \bea  \mathcal{L}_{SMT}&=& \mathcal{L}_{SMT}^R +
\mathcal{L}_{SMT}^{R\!\!\!\!/}, \eea where
 \bea -\mathcal{L}_{SMT}^R&=&
M^2_{ab}\widetilde{L}_{aL}^\dagger \widetilde{L}_{bL}+
m^2_{ab}\widetilde{l}_{aL}^{c *} \widetilde{l}_{bL}^c
+m^2_{Q1L}\widetilde{Q}_{1L}^\dagger \widetilde{Q}_{1
L}+m^2_{Q\alpha \bet L}\widetilde{Q}_{\alpha L}^\dagger
\widetilde{Q}_{\bet L}\crn &&+ m^2_{u_{ij}}\widetilde{u}_{iL}^{c
*} \widetilde{u}_{jL}^c + m^2_{d_{ij}}\widetilde{d}_{iL}^{c *}
\widetilde{d}_{jL}^c + m^2_{u^\prime }\widetilde{u^\prime}_{L}^{c
*} \widetilde{u^\prime}_{L}^c+ m^2_{d^\prime \al
\bet}\widetilde{d^\prime}_{\al L }^{c *} \widetilde{d^\prime}_{\bet
L}^c \crn && +\left\{ \eta_{ab}\widetilde{L}_{aL}\rho ^\prime
\widetilde{l}_{Lb}^c+ \varepsilon_{ab}\epsilon
\widetilde{L}_{aL}\widetilde{L}_{bL}\rho
+p\widetilde{Q}_{1L}\chi^\prime\widetilde{
u^\prime}^c_{L}\right.\crn && + p_{ \alpha i}\widetilde{Q}_{\alpha
L}\rho\widetilde{u}^c_{iL}+
h_{i}\widetilde{Q}_{1L}\rho^\prime\widetilde{d}^c_{iL}+h_{\al
i}\widetilde{Q}_{\alpha L}\chi\widetilde{d}^c_{iL} \crn &&+ \left.
p_{i}\widetilde{Q}_{1L}\chi^\prime\widetilde{u}^c_{iL} +
h^\prime_{\al}\widetilde{Q}_{1L}\rho^\prime\widetilde{d^\prime}^c_{\al
L} + h^\prime_{\alpha\beta}\widetilde{Q}_{\alpha
L}\chi\widetilde{d^\prime}^c_{\beta L}+r_{\al}\widetilde{Q}_{\alpha
L}\rho\widetilde{u^\prime}^c_{L}+ H.c.\right\}, \label{mmev}\eea and
\bea -\mathcal{L}_{SMT}^{R\!\!\!\!/} &=& M^{\prime 2}_a
\chi^\dagger\widetilde{L}_{aL}
 + \upsilon_{a}\epsilon\widetilde{L}_{aL}\chi\rho \crn
 && +
p_{5\alpha\beta\gamma}\widetilde{Q}_{\alpha L}
\widetilde{Q}_{{\beta} L}\widetilde{Q}_{{\gamma} L} +\kappa_{i\beta
j}\widetilde{d}_{iL}^c\widetilde{d^\prime}_{\beta
L}^c\widetilde{u}_{jL}^c+\vartheta_{i\beta
}\widetilde{d}_{iL}^c\widetilde{d^\prime}_{\beta
L}^c\widetilde{u^\prime}_{L}^c \nonumber \\ && +\pi_{i j
k}\widetilde{d}_{iL}^c\widetilde{d}_{j L}^c\widetilde{u}_{kL}^c
+\kappa_{4i k}\widetilde{d}_{iL}^c\widetilde{d}_{j
L}^c\widetilde{u^\prime}_{L}^c +\kappa_{5\alpha \beta
i}\widetilde{d^\prime}_{\alpha L}^c\widetilde{d^\prime}_{\beta
L}^c\widetilde{u}_{i L}^c \crn && + \kappa_{6\alpha \beta
}\widetilde{d^\prime}_{\alpha L}^c\widetilde{d^\prime}_{\beta
L}^c\widetilde{u^\prime}_{ L}^c + \om_{a\alpha j }\widetilde{L}_{a
L}\widetilde{Q}_{\alpha L}\widetilde{d}_{j L}^c+\om^\prime_{a\alpha
\beta }\widetilde{L}_{a L}\widetilde{Q}_{\alpha
L}\widetilde{d^\prime}_{\beta L}^c+ H.c.  \label{mme}\eea The
${R\!\!\!\!/}$ soft terms consist of  odd number of {\it
supersymmetric partners} - sfermions.

Note that  the last lines in (\ref{Rw1}) and (\ref{mmev}) contain
lepton-number violating terms (with $\De L = \pm 2)$. Hence we
have (see also \cite{dls1}) \bea \kappa_i, \vartheta'_\al,
\pi'_{\al}, \Pi'_{\al \bet}, p_i, r_\al, h'_{\al}, h'_{\al \bet}
\ll \kappa', \vartheta_i, \pi_{\al i}, \Pi_{\al i}, p, p_{\al i},
h_{i}, h_{\al i}. \label{lnvc} \eea

\section{\label{scalarpot}Scalar potential for sfermions}

The scalar potential of the model is a result of summation over $F$
and $D$ terms:
 \be V = F^{\phi *}F_\phi + \fr 1 2 \sum_a D^a D_a,
  \label{sf1}\ee where \cite{martin} \be F_\phi =
\fr{\pa W}{\pa \phi}, \hs W = W_2 + W_3,  \ee and \be D^a = -g
\left( \sum_\phi \phi^* T^a \phi \right). \label{sf1a}\ee The field
$\phi$ stands for all the scalars or sfermions in the model.

\subsection{F-term contribution}
\label{Fterm}

 From $W_2$ and $W_3$ we get
\bea F_{\chi^\prime} & = & \fr 1 2 (\mu_{0a} \widetilde{L}_{a L} +
\mu_\chi \chi) + \fr 1 3 (\kappa_{i} \tilde{Q}_{1L}
\tilde{u}^{c}_{iL} + \kappa^\prime \tilde{Q}_{1L}
\tilde{u}^{\prime c}_L ),\label{sf2}
\\
 F_{\chi^\si} & = & \fr 1 2 \mu_\chi  \chi^{\prime}_\si + \fr 1 3
\left(\la_{a} \epsilon_{m \si n} \tilde{L}_{a L}^m \rho^n +
\Pi_{\alpha i} \tilde{Q}_{\alpha L \si} \tilde{d}^{c}_{iL} +
\Pi^\prime_{\alpha \bet} \tilde{Q}_{\alpha L \si}
\tilde{d}^{\prime c}_{\bet L}\right),
 \label{sf4}\\
 F_{\rho^\si} & = &  \fr 1 2 \mu_\rho \rho^{\prime}_\si + \fr 1 3
\left(\la_{a} \epsilon_{m n \si } \tilde{L}_{a L}^m \chi^n  +
\la^\prime_{ab} \epsilon_{m n \si} \tilde{L}_{a L}^m \tilde{L}_{b
L}^n \right. \crn && \hs \hs + \left. \pi_{ \alpha i}
\tilde{Q}_{\alpha L \si}\tilde{u}^{c}_{iL} +\pi_{\alpha}^{\prime}
\tilde{Q}_{\alpha L \si}\tilde{u}^{\prime c}_{L} \right),
\label{sf7}
\\
 F_{\rho^\prime} & = & \fr 1 2 \mu_\rho \rho + \fr 1 3
\left(\ga_{ab}  \tilde{L}_{a L} \tilde{l}_{b L}^c +
\vartheta_{i}\tilde{Q}_{1L} \tilde{d}^{c}_{iL} +
\vartheta^\prime_{ \alpha}\tilde{Q}_{1L}  \tilde{d}^{\prime
c}_{\alpha L} \right), \label{sf9}\\
 F_{L_{aL}^\si} & = & \fr 1 2 \mu_{0a}  \chi^{\prime}_\si
 +\fr 1 3 (\ga_{ab}\rho'_\si \tilde{l}^c_{bL}
 +\la_{a} \epsilon_{\si m n} \chi^m \rho^n+2\la'_{ab} \epsilon_{ \si m n}
\tilde{L}_{b L}^m \rho^n\crn&&\hs \hs +\xi_{a\al j}\tilde{Q}_{\al
L\si}\tilde{d}^c_{jL} +\xi'_{a\al \beta}\tilde{Q}_{\al
L\si}\tilde{d}'^c_{\beta L})\\
F_{l^{c}_{L b}} & = &  \fr 1 3 \ga_{ab} \tilde{L}_{aL}
\rho^{\prime}, \label{sf13}\\
 F_{Q_{1 L}} & = & \fr 1 3 \left(\kappa_{i} \chi^\prime
\tilde{u}^{c}_{iL} + \kappa^\prime \chi^\prime
 \tilde{u}^{\prime c}_L  + \vartheta_{i} \rho^\prime
 \tilde{d}^{ c}_{iL}+ \vartheta_{ \al}^{\prime} \rho^\prime
 \tilde{d}^{\prime c}_{\al L}  \right),\label{sf30}
 \\
 F_{Q^\si_{\al  L}} & = &  \fr 1 3 \left(\pi_{ \al i}
 \rho_\si \tilde{u}^{c}_{iL}
+\pi_{\al}^{\prime} \rho_\si
 \tilde{u}^{\prime c}_L  + \Pi_{ \al i}
 \chi_\si \tilde{d}^{ c}_{iL} +
 \Pi_{ \al \bet}^{\prime} \chi_\si
 \tilde{d}^{\prime c}_{\bet L}  \right.\crn
&& + \left. 3 f_{\al \bet \ga} \ep_{\si j k} \widetilde{Q}^j_{\bet
L } \widetilde{Q}^k_{\ga L  } + \xi_{a \al i} \widetilde{L}_{a
L\si} \tilde{d}^c_{iL} + \xi'_{a \al\bet} \widetilde{L}_{a L\si}
\tilde{d'}^c_{\bet L}\right),
 \label{sf32}\\
 F_{u_{i L}^{c}} & = & \fr 1 3 \left( \kappa_{i} \chi^\prime
\widetilde{Q}_{1L} + \pi_{ \al i}  \rho \widetilde{Q}_{\al L}+
\xi_{1 j \bet i }\tilde{d}^c_{jL}\tilde{d'}^c_{\bet L}+ \xi_{3 k j
i }\tilde{d}^c_{k L}\tilde{d}^c_{j L}+ \xi_{5\al \bet  i
}\tilde{d'}^c_{\al L}\tilde{d'}^c_{\bet L} \right),\label{sf33a}
\\
   F_{u_{ L}^{\prime
c}} & = & \fr 1 3 \left( \kappa^\prime \chi^\prime
\widetilde{Q}_{1L} + \pi^\prime_{ \al}  \rho \widetilde{Q}_{\al L}
+ \xi_{2 i \bet }\tilde{d}^c_{iL}\tilde{d'}^c_{\bet L}+ \xi_{4 i j
 }\tilde{d}^c_{i L}\tilde{d}^c_{j L}+ \xi_{6\al \bet
}\tilde{d'}^c_{\al L}\tilde{d'}^c_{\bet L} \right),\label{sf33} \\
 F_{d_{i L}^ c} & = & \fr 1 3 \left( \vartheta_i\rho^\prime
\widetilde{Q}_{1L} + \Pi_{ \al i}  \chi \widetilde{Q}_{\al L}+
\xi_{1 i \bet j }\tilde{d'}^c_{\bet L}\tilde{u}^c_{jL} +\xi_{2 i
\bet }\tilde{d'}^c_{\bet L}\tilde{u'}^c_{L}\right.\crn&&\left.+ 2
\xi_{3 i j k }\tilde{d}^c_{j L}\tilde{u}^c_{k L}+2 \xi_{4 i j
 }\tilde{d}^c_{j L}\tilde{u'}^c_{ L}+ \xi_{a\al i
 }\tilde{L}_{a L}\tilde{Q}_{\al  L}\right),\label{sf3}
\\
 F_{d_{\alpha L}^{\prime c}} & = & \fr 1 3 \left(
\vartheta_\alpha^\prime\rho^\prime \widetilde{Q}_{1L} + \Pi_{\beta
\al }^\prime  \chi \widetilde{Q}_{\beta L} + \xi_{1 i \al j
}\tilde{d}^c_{i L}\tilde{u}^c_{jL} +\xi_{2 i \al }\tilde{d}^c_{i
L}\tilde{u'}^c_{L}\right.\crn&&\left.+ 2 \xi_{5\al \bet
}\tilde{d'}^c_{\bet L}\tilde{u}^c_{i L}+ 2 \xi_{6\al \bet
}\tilde{d'}^c_{\bet L}\tilde{u'}^c_{ L}+ \xi'_{ a \bet \al
 }\tilde{L}_{a L}\tilde{Q}_{\bet  L}\right).\label{sf5}
 \eea With these $F$-terms,
besides the second order  mass terms in $V$, we also get trilinear
and quartic couplings  of the sfermions. Below only the mass terms
and the linear (by fields) terms   are our interest.

\subsection{D-term contribution}
\label{Dterm}

By  Eq. (\ref{sf1a}), we separate two subgroups, namely $SU(3)_L$
and $U(1)_X$. \ben
\item {\it $D$-term contribution from  $SU(3)_L$}:

The interested contribution to sfermion masses has a form \bea D^a
= -g \left[\sum_{sfermions} \tilde{f}^\dag T^a \tilde{f} +
\sum_{Higgs} H^\dag T^a H \right].\label{sf181} \eea Since $T_a =
T_a^\dag$, we have \bea (D^a)^* D_a = 2
g^2\left[\left(\sum_{sfermions} \tilde{f}^\dag T^a
\tilde{f}\right) \left(\sum_{Higgs} H^\dag T^a H \right)\right] +
\cdot \cdot \cdot, \label{sf182} \eea where $\cdot \cdot \cdot$
are the terms which do not contribute to sfermion masses.  The
factor 2 in (\ref{sf182}) is the Newton's binomial coefficient.
Since sfermion masses are our interest, therefore, in the second
factor in (\ref{sf182}), only the diagonal $T_3$ , $T_8$ and
non-diagonal $T_4$ satisfy this purpose. Let us calculate the
second factor in (\ref{sf182}): \bea H_3 & \equiv & \sum_{H= \chi,
\chi^\prime, \rho, \rho^\prime } <H^\dag > T_3 < H
> = \fr 1 4 (u^2 - u^{\prime 2} ) - \fr 1 4 (v^2 - v^{\prime 2})\crn
& = & -\fr 1 4 \left(u^2 \fr{\cos 2\bet}{s_\bet^2} + v^2 \fr{\cos
2\ga}{c_\ga^2} \right), \label{sf184}\\ H_8 & \equiv & \sum_{H=
\chi, \chi^\prime, \rho, \rho^\prime } <H^\dag > T_8 < H
>\crn
 &=& \fr{1}{2\sqrt{3}} \left\{ \fr 1 2 (u^2 - u^{\prime
2} ) +  \fr 1 2 (v^2 - v^{\prime 2} ) -  (w^2 - w^{\prime 2}
)\right\} \crn & = &   \fr{1}{4\sqrt{3}} \left[ v^2 \fr{\cos
2\ga}{c_\ga^2}- (u^2 - 2 w^2)\fr{\cos 2\bet}{s_\bet^2}
\right]\label{sf185}\\
H_4 & \equiv & \sum_{H= \chi, \chi^\prime, \rho, \rho^\prime }
<H^\dag > T_4 < H>\crn &=&\fr 1 2 (u w-u' w')=-\fr 1 2 u w \fr{\cos
2\beta}{s^2_\beta}.\label{sf199} \eea In (\ref{sf184}),
(\ref{sf185}) and (\ref{sf199}) we have used \cite{susyeco,{higph}}
\be \tan \bet = \fr{u}{u^\prime} = \fr{w}{w^\prime}, \hs \tan \ga =
\fr{v^\prime}{v}.\ee Here we have taken into account that for
antitriplets, $T_a, a= 3,8,4$  changes a sign. Note that the
contribution from $T_4$ is proportional to $u$ -- the lepton number
violating parameter.

Let us  consider the first factor in (\ref{sf182}). Since the
singlet fields do not give contribution, hence  for sleptons we
have: \bea SL_3 & \equiv & \tilde{L}_{ a L} ^\dag T_3 \tilde{L}_{ a
L}  = \fr 1 2 \tilde{\nu}^*_{ a L} \tilde{\nu}_{ a L}  - \fr 1 2
\tilde{l}^*_{a L}\tilde{l}_{ a L} ,\label{sf186}\\
 SL_8  & \equiv  & \tilde{L}_{ a L}^\dag
T_8 \tilde{L}_{ a L}  =  \fr{1}{\sqrt{3}}\left(\fr 1 2 \tilde{\nu}_{
a L}^* \tilde{\nu}_{ a L} + \fr 1 2 \tilde{l}^*_{ a L}\tilde{l}_{ a
L} - \tilde{\nu}_{ a L}^{c *}\tilde{\nu}_{ a L}^{c
}\right),\label{sf187}\\
 SL_4 & \equiv & \tilde{L}_{ a L} ^\dag T_4 \tilde{L}_{ a L}  =
\fr 1 2 \tilde{\nu}^*_{ a L} \tilde{\nu}^c_{ a L}  + \fr 1 2
\tilde{\nu}^{c*}_{a L}\tilde{\nu}_{ a L}.\eea Analogously for
squarks, the contributions from one triplet and two antitriplets
are: from the first triplet
 \bea SQ_3 & \equiv & \widetilde{Q}_{1 L}^\dag T_3
\widetilde{Q}_{1 L} = \fr 1 2 \tilde{u}_{1 L}^* \tilde{u}_{1 L} -
\fr 1 2 \tilde{d}_{1 L}^*
\tilde{d}_{1 L},\label{sf188}\\
 SQ_8 & \equiv & \widetilde{Q}_{1
L}^\dag T_8 \widetilde{Q}_{1 L} = \fr{1}{\sqrt{3}}\left(\fr 1 2
\tilde{u}_{1 L}^* \tilde{u}_{1 L} + \fr 1 2  \tilde{d}_{1 L}^*
\tilde{d}_{1 L} - \tilde{u}_{ L}^{\prime *} \tilde{u}_{ L}^\prime
\right),\label{sf189}\\
SQ_4 & \equiv & \widetilde{Q}_{1 L}^\dag T_4 \widetilde{Q}_{1 L} =
\fr 1 2 \tilde{u}_{1 L}^* \tilde{u}^\prime_{L} + \fr 1 2
\tilde{u}^{\prime *}_{L} \tilde{u}_{1 L},\eea
 from two  antitriplets: \bea SaQ_3 & \equiv & -
\widetilde{Q}_{\al L}^\dag T_3 \widetilde{Q}_{\al L} = - \fr 1 2
\tilde{d}_{\al L}^* \tilde{d}_{\al L} + \fr 1 2 \tilde{u}_{\al L}^*
\tilde{u}_{\al L} ,\label{sf190}\\
 SaQ_8 & \equiv & -\widetilde{Q}_{\al
L}^\dag T_8 \widetilde{Q}_{\al  L} =  - \fr{1}{2 \sqrt{3}}\left(
\tilde{u}_{\al L}^* \tilde{u}_{\al L} + \tilde{d}_{\al L}^*
\tilde{d}_{\al L} - 2 \tilde{d}_{\al L}^{\prime
*} \tilde{d}_{\al L}^\prime \right),\label{sf191}\\
SaQ_4 & \equiv & - \widetilde{Q}_{\al L}^\dag T_4
\widetilde{Q}_{\al L} = - \fr 1 2 \tilde{d}_{\al L}^*
\tilde{d}^\prime_{\al L} - \fr 1 2 \tilde{d}_{\al L}^{\prime *}
\tilde{d}_{\al L}.\eea Thus, the contribution from $SU(3)_L$
subgroup to slepton masses are: \be g^2 ( SL_3 \times H_3 + SL_8
\times H_8+SL_4\times H_4), \label{sf192}\ee and to squark masses:
 \be    g^2[(SQ_3 + SaQ_3) \times H_3 + (SQ_8 + SaQ_8)
\times H_8+(SQ_4 + SaQ_4) \times H_4]. \label{sf193}\ee

\item {\it $D$-term contribution from  $U(1)_X$}:

First, for the Higgs part, we have \bea H_1 & \equiv  &\sum_{H=
\chi, \chi^\prime, \rho, \rho^\prime } <H^\dag > X < H
> \crn & = & - \fr 1 6 (u^2 -
u^{\prime 2} ) + \fr 2 6  (v^2 - v^{\prime 2}) - \fr 1 6 (w^2 -
w^{\prime 2} ) \crn & = &  \fr 1 6 \left[(u^2 +  w^2)\fr{\cos
2\bet}{s_\bet^2} + 2 v^2 \fr{\cos 2\ga}{c_\ga^2} \right]
 \label{sf194}\eea Similarly, for sleptons
\bea SL_1 & \equiv  & - \fr 1 3 (\tilde{\nu}^*_{ a L} \tilde{\nu}_{
a L} + \tilde{l}^*_{ a L}\tilde{l}_{ a L} + \tilde{\nu}^{c*}_{ a L}
\tilde{\nu}^c_{ a L} ) + \tilde{l}^{c *}_{ a L}\tilde{l}^c_{ a
L}.\label{sf195}\eea For squarks in the first generation we get:
\bea SQ_1 & \equiv  & \fr 1 3 (\tilde{u}_{1 L}^* \tilde{u}_{1 L} +
\tilde{d}_{1 L}^* \tilde{d}_{1 L}
  + \tilde{u}_{ L}^{\prime *}
\tilde{u}_{ L}^\prime  ) - \fr 2 3 (\tilde{u}_{1 L}^{c *}
\tilde{u}^c_{1 L} + \tilde{u}_{ L}^{ \prime c *} \tilde{u}^{ \prime
c}_{ L} ) + \fr 1 3 \tilde{d}_{1 L}^{c *} \tilde{d}^c_{1 L}.
 \label{sf196}\eea For squarks in the last two generations we get
 also: \bea SaQ_1  & \equiv  &  - \fr 2 3 \tilde{u}_{\al L}^{c *}
\tilde{u}^c_{\al L} + \fr 1 3 (\tilde{d}_{\al L}^{c *}
\tilde{d}^c_{\al  L} + \tilde{d}_{\al L}^{ \prime c *} \tilde{d}^{
\prime c}_{\al L}).
 \label{sf196}\eea
The contribution from subgroup  $U(1)_X$ to slepton masses is \be
  g^{\prime 2}\times SL_1 \times H_1 =  g^2 t^2 \times SL_1 \times H_1
 \label{sf197}\ee where \cite{dl}
 \be t^2 = (g^\prime / g)^2 = \fr{3 s_W^2}{3 - 4 s_W^2}\ee
  and to squark masses: \be
  g^{\prime 2}\times H_1
(SQ_1 + SaQ_1) =  g^2 t^2 \times  H_1 (SQ_1 + SaQ_1).
\label{sf198}\ee
 \een

  The total contribution is a result of summation over
two above mentioned subgroup parts. In contradiction to the MSSM,
the contribution from $T_4$ is lepton number violating ($\De L =
\pm 2 $). We will deal with this in next section. It is easy to
realize that {\it the $D$-term contributions are diagonal}.

 \section{\label{slepm}Slepton masses}

 Relevant mass terms for  sleptons arisen from the $F, D$-terms and
 the soft terms are
 as follows:
 \bea \mathcal{L}_{slepton} & = & M^2_{ab}
 \widetilde{L}_{aL}^* \widetilde{L}_{bL}+
m^2_{ab}\widetilde{l}_{aL}^{c *} \widetilde{l}_{bL}^c+  \fr 1 4
\mu_{0a} \mu_{0b} \tilde{L}_{aL}^* \tilde{L}_{bL} \crn &&+\left[
 M^{\prime 2}_a
\chi^*\widetilde{L}_{aL}+\eta_{ab}\widetilde{L}_{aL}\rho ^\prime
\widetilde{l}_{Lb}^c\right.\crn &&
+\upsilon_{a}\epsilon\widetilde{L}_{aL}\chi\rho+
\varepsilon_{ab}\epsilon \widetilde{L}_{aL}\widetilde{L}_{bL}\rho
\crn && + \fr 1 4 \mu_{0a} \mu_{\chi}\chi^* \tilde{L}_{aL}  + \fr 1
6 \mu_\chi \la_{a} \epsilon \tilde{L}_{a L}\chi^{\prime *} \rho \crn
&&+ \fr 1 6 \mu_\rho \left( \la_{a} \epsilon \tilde{L}_{a L} \chi
\rho^{\prime *} + \la^\prime_{ab} \epsilon \tilde{L}_{a L}
\tilde{L}_{b L} \rho^{\prime *}
 \right)+\fr 1 6 \mu_\rho \rho^* \left(\ga_{ab}
\tilde{L}_{a L} \tilde{l}_{b L}^c\right) \crn & & + \fr 1 6
\mu_{0a} \left(\ga_{ab} \chi^{\prime  *}.\rho^{\prime}
\tilde{l}_{b L}^c + 2 \la^\prime_{ab} \epsilon \chi^{\prime
*}\tilde{L}_{bL} \rho \right) + \fr 1 9 \ga_{ab}\la_{a}
\epsilon\rho^{\prime} \chi^*\rho^*.\tilde{l}_{b L}^c \crn
&&\left.+  \fr 2 9 \la^\prime_{ab}\la_a[(\chi^*\tilde{L}_{b
L})(\rho^*\rho)-(\rho^*\tilde{L}_{b
L})(\chi^*\rho)]+H.c\right]\crn && + \fr 1 9 \ga_{ab}\ga_{ab'}
\rho^{\prime*}\rho' \tilde{l}_{b L}^c\tilde{l}_{b' L}^{c*}+ \fr 1
9 \ga_{ab} \ga_{a'b} (\tilde{L}_{aL} \rho^{\prime})
(\tilde{L}_{a'L} \rho^{\prime})^* \crn &&+ \fr1 9 \la_a \la_b
[(\tilde{L}^*_{aL}\tilde{L}_{bL})(\rho^*\rho)
-(\tilde{L}^*_{aL}\rho)(\rho^*\tilde{L}_{bL})]\crn &&+ \fr1 9
\la_a \la_b [(\tilde{L}^*_{aL}\tilde{L}_{bL})(\chi^*\chi)
-(\tilde{L}^*_{aL}\chi)(\chi^*\tilde{L}_{bL})]\crn &&+\fr 4 9
\la^\prime_{ca}\la^\prime_{cb}[(\tilde{L}^*_{aL}\tilde{L}_{bL})(\rho^*\rho)
-(\tilde{L}^*_{aL}\rho)(\rho^*\tilde{L}_{bL})]\crn &&  +   g^2 (
SL_3 \times H_3 + SL_8 \times H_8 + SL_4 \times H_4)+   g^2 t^2
\times SL_1 \times H_1. \label{sf15}\eea

Expanding the $D$-term  contribution
  [in the last line of (\ref{sf15})] yields
 \bea
  D_L & \equiv &  g^2 ( SL_3 \times  H_3 + SL_8 \times
H_8 + SL_4 \times  H_4) +   g^2 t^2 \times SL_1 \times H_1\crn & =
&   g^2 \left\{ \tilde{\nu}^*_{ a L} \tilde{\nu}_{ a L} \left(\fr
1 2 H_3 + \fr{1}{2 \sqrt{3}}H_8 - \fr{t^2}{3} H_1\right)\right.
\crn && \hs + \tilde{l}^*_{ a L} \tilde{l}_{ a L} \left(- \fr 1 2
H_3 + \fr{1}{2 \sqrt{3}}H_8 - \fr{t^2}{3} H_1\right)\crn &&  \hs
\hs  + \left. \tilde{\nu}_{ a L}^{c *}\tilde{\nu}_{ a L}^{c }
\left( - \fr{1}{ \sqrt{3}}H_8 - \fr{t^2}{3} H_1\right)  +
\tilde{l}^{c *}_{ a L}\tilde{l}^c_{ a L} t^2 H_1 + \fr 1 2
(\tilde{\nu}^*_{ a L} \tilde{\nu}^c_{ a L} H_4 + H.c.) \right\}
\label{sf601}\eea

The terms containing mixture of  scalar Higgs bosons among sleptons
followed from $\mathcal{L}_{slepton}$ is:
 \bea \mathcal{L}_{mix} & = &
 M^{\prime 2}_a
\chi^*\widetilde{L}_{aL}
+\upsilon_{a}\epsilon\widetilde{L}_{aL}\chi\rho \crn && + \fr 1 4
\mu_{0a} \mu_{\chi}\chi^* \tilde{L}_{aL}  + \fr 1 6 \mu_\chi \la_{a}
\epsilon \tilde{L}_{a L}\chi^{\prime *} \rho \crn &&+ \fr 1 6
\mu_\rho \la_{a} \epsilon \tilde{L}_{a L} \chi \rho^{\prime
*} + \fr 1 9 \ga_{ab}\la_{a} \epsilon\rho^{\prime}
\chi^*\rho^*.\tilde{l}_{b L}^c\crn & & + \fr 1 6 \mu_{0a}
\left(\ga_{ab} \chi^{\prime  *}.\rho^{\prime} \tilde{l}_{b L}^c + 2
\la^\prime_{ab} \epsilon \chi^{\prime *}\tilde{L}_{bL} \rho \right)
 \crn &&+  \fr 2 9
\la^\prime_{ab}\la_a[(\chi^*\tilde{L}_{b
L})(\rho^*\rho)-(\rho^*\tilde{L}_{b L})(\chi^*\rho)]+H.c.
\crn &&=\left(M^{\prime 2}_a+\fr 1 4 \mu_{0a} \mu_{\chi}\right)
(\chi^{0*}_1\widetilde{\nu}_{aL}+\chi^+\widetilde{l}_{aL}
+\chi^{0*}_2\widetilde{\nu}^c_{aL}) \crn
&&+\upsilon_{a}[(-\widetilde{\nu}_{aL}\chi^0_2
+\widetilde{\nu}^c_{aL}\chi^0_1)\rho^0
+\widetilde{l}_{aL}(-\chi^0_1\rho_2^+ +\chi^0_2\rho_1^+) ]\crn &&
+\left(\fr 1 6 \mu_\chi \la_{a}- \fr 1 3 \mu_{0b}
\la^\prime_{ba}\right) [(-\widetilde{\nu}_{aL}\chi'^{0*}_2
+\widetilde{\nu}^c_{aL}\chi'^{0*}_1)\rho^0
+\widetilde{l}_{aL}(-\chi'^{0*}_1\rho_2^+ +\chi'^{0*}_2\rho_1^+)
]\crn &&+ \fr 1 6 \mu_\rho \la_{a}[(-\widetilde{\nu}_{aL}\chi^0_2
+\widetilde{\nu}^c_{aL}\chi^0_1)\rho'^{0*}
+\widetilde{l}_{aL}(-\chi^0_1\rho'^+_2 +\chi^0_2\rho'^+_1) ]\crn &&+
\fr 1 9 \ga_{ab}\la_{a} [(-\rho^{\prime -}_1
\chi^{0*}_2+\rho^{\prime -}_2 \chi^{0*}_1)\rho^{0*}+\rho^{\prime 0}
(\chi^{0*}_2\rho^{-}_1-\chi^{0*}_1\rho^{-}_2)]\tilde{l}_{b L}^c\crn
& & + \fr 1 6 \mu_{0a} \ga_{ab}(\chi'^{0*}_1 \rho'^-_1+ \chi^{\prime
-}.\rho^{\prime 0}+\chi'^{0*}_2 \rho'^-_2) \tilde{l}_{b L}^c
 \crn &&+  \fr 2 9
\la^\prime_{ab}\la_a[(\chi^{0*}_1\tilde{\nu}_{b
L}+\chi^+\tilde{l}_{b L}+\chi^{0*}_2\tilde{\nu}^c_{b
L})(\rho^{0*}\rho^0)\crn &&-\rho^{0*}\tilde{l}_{b
L}(\chi^{0*}_1\rho^+_1+\chi^+\rho^0+\chi^{0*}_2\rho^+_2)]+H.c.
\label{thuy1} \eea

Now we have to expand neutral Higgs fields around the VEVs as \bea
\chi^T&=&\left(
           \begin{array}{ccc}
\fr{u+S_1+iA_1}{\sqrt{2}}, & \chi^{-}, & \fr{w+S_2+iA_2}{\sqrt{2}} \\
           \end{array}
         \right), \hs  \rho^T = \left(
\begin{array}{ccc}
  \rho_1^+, & \fr{v+S_5+iA_5}{\sqrt{2}}, & \rho_2^+ \crn
          \end{array}
      \right),\label{2}\\ {\chi^\prime}^T&=&\left(
           \begin{array}{ccc}
\fr{u^\prime+S_3+iA_3}{\sqrt{2}},
& \chi^{\prime +}, & \fr{w^\prime+S_4+iA_4}{\sqrt{2}} \\
           \end{array}
         \right),\hs {\rho ^\prime}^T =\left(
                         \begin{array}{ccc}
                           \rho_1^{\prime -}, &
\fr{ v^\prime +S_6+iA_6}{\sqrt{2}}, & \rho_2^{\prime-} \\
                         \end{array}\right).\label{sf17}\eea
where for short, the neutral scalar is expressed through the VEV and
physical field (in breve) as follows \be h^0 = \fr{1}{\sqrt{2}}(vev
+ \breve{h}). \label{sf19}\ee

From  Eq. (\ref{thuy1}) we see that, there is mixing among charged
Higgs boson $\chi^{\prime -}$ with $ \tilde{l}_{b L}^c $ as well as
neutral Higgs fields $\breve{\chi}_1^{\prime 0 }$ with neutral
sleptons such as $\tilde{\nu}^c_{bL}$, etc.  To remove this mixing,
we have to impose $R$-parity condition.

Imposing $R$-parity conservation on  (\ref{thuy1}) yields
 \bea && M^{\prime 2}_a +
\fr 1 4 \mu_{0a} \mu_{\chi} = 0,\hs \upsilon_{a}=0,\hs \mu_\rho
\la_{a} =0,\label{thuy912a}\\ &&  \fr 1 6 \mu_\chi \la_{a}-\fr 1 3
\mu_{0b} \la^\prime_{ba}=0,\hs \ga_{ab}\la_a=0,\hs
\la'_{ab}\la_a=0,
\label{sf23}\\
&& \mu_{0a} \ga_{ab} = 0.\label{thuy2} \eea

Note that the conditions in (\ref{thuy912a})--(\ref{thuy2}) contain
also the constraint equations at the tree level for $\tilde{\nu}_{a
L}$ and $\tilde{\nu}^c_{a L}$.

 Taking into account of (\ref{thuy912a})--(\ref{thuy2}),
the slepton mass Lagrangian becomes
\bea \mathcal{L}_{slepton} & = &  D_L + \left(M^2_{ab} + \fr 1 4
\mu_{0a} \mu_{0b}\right)(\tilde{\nu}_{a L}^* \tilde{\nu}_{b L} +
\tilde{l}_{a L}^* \tilde{l}_{b L} + \tilde{\nu}_{a L}^{c*}
\tilde{\nu}^c_{b L}) + m^2_{ab}\widetilde{l}_{aL}^{c *}
\widetilde{l}_{bL}^c \crn && +\left\{\fr{1}{\sqrt{2}}
\left(\eta_{ab} v^\prime + \fr 1 6 \mu_\rho
\ga_{ab}v\right)\widetilde{l}_{aL} \widetilde{l}_{bL}^c -
\fr{1}{\sqrt{2}}\varepsilon_{ab} v
(\tilde{\nu}_{aL}\tilde{\nu}^c_{bL}
-\tilde{\nu}_{bL}\tilde{\nu}^c_{aL})\right. \crn && - \left.
\fr{1}{6 \sqrt{2}}\mu_\rho  \la^\prime_{ab}v' ( \tilde{\nu}_{a L}
\tilde{\nu}^c_{b L} -\tilde{\nu}_{b L} \tilde{\nu}^c_{a L})
  + H.c.\right\} \crn &&
 + \fr 1 9 \ga_{ab}
\ga_{ab'} \tilde{l}_{b L}^{c *} \tilde{l}_{b' L}^c \fr{v^{\prime
2}}{2} + \fr 1 9 \ga_{ab} \ga_{a'b} (\tilde{l}_{aL}
\tilde{l}^\dag_{a'L})\fr{v^{\prime
2}}{2}\crn&&+\fr{1}{18}v^2(\la_a \la_b+4
\la^\prime_{ca}\la^\prime_{cb})(\tilde{\nu}^*_{aL}\tilde{\nu}_{bL}
+\tilde{\nu}^{c*}_{aL}\tilde{\nu}^c_{bL})\crn && + \fr{1}{18}\la_a
\la_b \left[w^2 \tilde{\nu}^*_{aL}\tilde{\nu}_{bL} +u^2
\tilde{\nu}^{c*}_{aL}\tilde{\nu}^c_{bL} \right. \crn
&&\left.+(u^2+w^2)\tilde{l}^*_{aL}\tilde{l}_{bL} -u
w(\tilde{\nu}^*_{aL}\tilde{\nu}_{bL}^c
+\tilde{\nu}^{c*}_{aL}\tilde{\nu}_{bL})\right]. \label{sf24}\eea

\subsection{\label{cslepm}Charged sleptons}

From  Eq. (\ref{sf24}), the mass Lagrangian for charged sleptons
is given by
 \bea \mathcal{L}_{Charlepton} & = & \left[M^2_{ab} +
\fr 1 4 \mu_{0a} \mu_{0b} +  \fr{
 v^{\prime 2}}{18}\ga_{ca} \ga_{cb}
 + \fr{g^2}{2} \de_{a b}\left(- H_3
 + \fr{1}{\sqrt{3}}H_8 - \fr{2 t^2}{3}H_1 \right) \right]
 \tilde{l}_{a L}^* \tilde{l}_{b L} \crn && + \left(m^2_{ab}
+ \fr{ v^{\prime 2}}{18}\ga_{ca}
\ga_{cb} +  g^2 t^2 H_1 \de_{a
b}\right)
 \widetilde{l}_{aL}^{c *}
\widetilde{l}_{bL}^c \crn && +\left[ \fr{1}{\sqrt{2}}
\left(\eta_{ab} v^\prime + \fr 1 6 \mu_\rho
\ga_{ab}v\right)\widetilde{l}_{aL} \widetilde{l}_{bL}^c +
H.c.\right]\crn && + \fr{1}{18}\la_a \la_b
(u^2+w^2)\tilde{l}^*_{aL}\tilde{l}_{bL}  \label{sf25}\eea
 For analysis below, let us denote
 \bea B_{ab} & = & M^2_{ab} + \fr 1
4 \mu_{0a} \mu_{0b} + \fr{
 v^{\prime 2}}{18}\ga_{ca} \ga_{cb}
+ \fr{1}{18}\la_a \la_b (u^2+w^2)\crn && + \fr{g^2}{2} \de_{a
b}\left(- H_3
 + \fr{1}{\sqrt{3}}H_8 - \fr{2 t^2}{3}H_1 \right),\\
 C_{ab}& =  &m^2_{ab} + \fr{
 v^{\prime 2}}{18}\ga_{ca} \ga_{cb} +  g^2 t^2 H_1 \de_{a b}
,\\
 D_{ab} &= &\fr{1}{\sqrt{2}} \left(\eta_{ab} v^\prime + \fr 1 6
\mu_\rho \ga_{ab}v\right).
 \label{sf26}\eea
For the sake of convenience, let us denote $ \widetilde{l}_{aL}^{c
*} \equiv \widetilde{l}_{a R}$.
Then, in the base $(\tilde{l}_{a L},  \tilde{l}_{bR})$ =
 $(\tilde{l}_{1 L}$, $ \tilde{l}_{2L}$, $\tilde{l}_{3L}$,
 $\tilde{l}_{1 R}$, $\tilde{l}_{2 R}$, $ \tilde{l}_{3 R})$, the mass
 matrix is given by
 \bea
\left(%
\begin{array}{cc}
B_{ab} & D_{ab}\\
D_{ab} & C_{ab}
\end{array}%
\right).
 \label{sf27} \eea
To deal with this $ 6 \times 6$ matrix, following
Ref.\cite{martin}, we assume that there is substantial  mixing
among $(\tilde{\tau}_L, \tilde{\tau}_R)$ {\it only}. Hereafter, we
adopt $\tilde{\tau} = \tilde{l}_3$, $\tilde{t} = \tilde{u}_3$,
$\tilde{b} = \tilde{d}_3$, etc. This means that, non-vanishing
matrix elements in (\ref{sf27}) are $ B_{11} $, $B_{22}$, $
B_{33}$, $C_{11}$, $ C_{22}$,  $ C_{33}$, $ D_{33}$.

Diagonalizing the above matrix, we get eigenmasses and eigenstates
given in Table~\ref{eigen}.
\begin{table}[h]
\caption{Masses and eigenstates of charged sleptons}
\begin{center}
\begin{tabular}{|c|c|c|c|c|}
\hline
 Eigenstate &$\tilde{l}_{1 L}$ & $\tilde{l}_{2 L}$ &
$\tilde{l}_{1 R}$ & $\tilde{l}_{2 R}$
\\
\hline
 $(\textrm{Mass})^2$ &$B_{11}$ & $B_{22}$ & $C_{11}$ & $C_{22}$   \\
 \hline
\end{tabular}
\label{eigen}
\end{center}
\end{table}
and two  others are
 \bea \tilde{\tau}_{L} &=&
s_{\theta_s}\tilde{l}_{3 R}
- c_{\theta_s}\tilde{l}_{3 L}, \label{sf27a}\\
\tilde{\tau}_{R} &=& c_{\theta_s}\tilde{l}_{3
R}+s_{\theta_s}\tilde{l}_{3 L}, \label{sf27b}\eea with respective
masses \bea
m^2_{\tilde{\tau}_{L}}&=&  \fr 1 2 (B_{33} + C_{33} - \De),\\
m^2_{\tilde{\tau}_{R}}&=&  \fr 1 2 (B_{33} + C_{33} + \De),
\label{sf27c}\eea where \bea \De &=& \sqrt{(C_{33}- B_{33})^2 +
4D_{33}^2},\label{sf35}\\
t_{2\theta_s}&=&\fr{2D_{33}}{C_{33}-B_{33}}.\label{sf36}\eea
 With the mentioned assumption, we have \bea &&
  m^2_{\tilde{l}_{1 L}}=M^2_{11} + \fr 1 4
\mu_{01}^2 + \fr{
 v^{\prime 2}}{18}\ga_{c1}^2 +
 \fr{1}{18}\la_1^2(u^2+w^2)- \fr{g^2}{2} \left( H_3
 - \fr{1}{\sqrt{3}}H_8 + \fr{2 t^2}{3}H_1 \right),\label{sf37}\\
&&m^2_{\tilde{l}_{2 L}}= M^2_{22} + \fr 1 4 \mu_{02}^2 +  \fr{
 v^{\prime 2}}{18}\ga_{c2}^2 + \fr{1}{18}
 \la_2^2  (u^2+w^2)- \fr{g^2}{2} \left( H_3
 - \fr{1}{\sqrt{3}}H_8 + \fr{2 t^2}{3}H_1 \right),\label{sf39}\\
 &&   m^2_{\tilde{l}_{1R}} = m^2_{11} + \fr{
 v^{\prime 2}}{18}\ga_{c1}^2  +  g^2 t^2 H_1,
 \label{sf40}\\
&&  m^2_{\tilde{l}_{2R}}= m^2_{22} + \fr{
 v^{\prime 2}}{18}\ga_{c2}^2 +  g^2 t^2 H_1,\label{sf41}\\
\label{sf42} \eea
 For the highest sleptons -  staus:
 \bea  m^2_{\tilde{\tau}_{ L}} & = & \fr {1}
{2}\left[M_{33}^2+m_{33}^2+\fr{v^{\prime 2}}{9}\ga_{c3}^2+\fr {1}{4}
\mu_{03}^{2} + \fr{1}{18}\la_3^2 (u^2+w^2)\right.\crn&& \left.  -
\fr{g^2}{2} \left( H_3
 - \fr{1}{\sqrt{3}}H_8 - \fr{4 t^2}{3}H_1 \right)
-\De\right],\label{sf44}\\
 m^2_{\tilde{\tau}_{ R}} & = & \fr {1}
{2}\left[M_{33}^2+m_{33}^2+\fr{v^{\prime 2}}{9}\ga_{c3}^2+\fr {1}{4}
\mu_{03}^{2} + \fr{1}{18}\la_3^2(u^2+w^2)\right.\crn&& \left.-
\fr{g^2}{2} \left( H_3
 - \fr{1}{\sqrt{3}}H_8 - \fr{4 t^2}{3}H_1 \right)
+\De\right],\label{sf45} \eea with \bea \De &=&
\left\{\left[M_{33}^2 + \fr 1 4 \mu_{03}^2 + \fr{1}{18}\la_3^2
(u^2+w^2)-m_{33}^2\right.\right.\crn&& \left.\left. - \fr{g^2}{2}
\left( H_3
 - \fr{1}{\sqrt{3}}H_8 + \fr{8 t^2}{3}H_1 \right)\right]^2 +
2 \left(\eta_{33} v^\prime + \fr 1 6 \mu_\rho
\ga_{33}v\right)^2\right\}^\fr 1 2 \label{sf46}\eea

\subsection{\label{nlepm}Sneutrinos}

 Eq. (\ref{sf24}) provides the following
 mass Lagrangian for sneutrinos:
\bea \mathcal{L}_{sneutrinos} & = &\left[ \fr{g^2}{2} \de_{a
b}\left( H_3
 + \fr{1}{\sqrt{3}}H_8 - \fr{2 t^2}{3}H_1 \right) + M^2_{ab} + \fr 1 4
\mu_{0a} \mu_{0b}\right]\tilde{\nu}_{a L}^* \tilde{\nu}_{b L} \crn
&& + \left[- g^2 \de_{a b}\left(
  \fr{1}{\sqrt{3}}H_8 + \fr{ t^2}{3}H_1 \right) +
  M^2_{ab} + \fr 1 4 \mu_{0a} \mu_{0b}\right]
\tilde{\nu}_{a L}^{c*} \tilde{\nu}^c_{b L}  \crn &&  -\left[
\fr{1}{\sqrt{2}}\varepsilon_{ab} v
(\tilde{\nu}_{aL}\tilde{\nu}^c_{bL}
-\tilde{\nu}_{bL}\tilde{\nu}^c_{aL})\right.\crn &&\left.+ \fr{1}{6
\sqrt{2}}\mu_\rho \la^\prime_{ab}v' ( \tilde{\nu}_{a L}
\tilde{\nu}^c_{b L} -\tilde{\nu}_{b L} \tilde{\nu}^c_{a L})
  + H.c. \right]\crn &&+\fr{1}{18}v^2(\la_a \la_b+4
\la^\prime_{ca}\la^\prime_{cb})(\tilde{\nu}^*_{aL}\tilde{\nu}_{bL}
+\tilde{\nu}^{c*}_{aL}\tilde{\nu}^c_{bL})\crn && + \fr{1}{18}\la_a
\la_b \left[w^2 \tilde{\nu}^*_{aL}\tilde{\nu}_{bL} +u^2
\tilde{\nu}^{c*}_{aL}\tilde{\nu}^c_{bL}
 -u w(\tilde{\nu}^*_{aL}\tilde{\nu}_{bL}^c
+\tilde{\nu}^{c*}_{aL}\tilde{\nu}_{bL})\right] \crn &&- \fr{g^2}{4}
u w \fr{cos 2\beta}{s^2_\beta}(\tilde{\nu}^*_{ a L} \tilde{\nu}^c_{
a L} + H.c.)\crn
    & = &
\left[ \fr{g^2}{2} \de_{a b}\left( H_3
 + \fr{1}{\sqrt{3}}H_8 - \fr{2 t^2}{3}H_1 \right) + M^2_{ab} + \fr 1 4
\mu_{0a} \mu_{0b}\right.\crn &&\left.+\fr{1}{18}v^2(\la_a \la_b+4
\la^\prime_{ca}\la^\prime_{cb})+ \fr{1}{18}\la_a \la_b w
^2\right]\tilde{\nu}_{a L}^* \tilde{\nu}_{b L} \crn && + \left[-
g^2 \de_{a b}\left(
  \fr{1}{\sqrt{3}}H_8 + \fr{ t^2}{3}H_1 \right) +
  M^2_{ab} + \fr 1 4 \mu_{0a} \mu_{0b}\right.\crn
  &&\left.+\fr{1}{18}v^2(\la_a \la_b+4
\la^\prime_{ca}\la^\prime_{cb})+ \fr{1}{18}\la_a \la_b  u^2\right]
\tilde{\nu}_{a L}^{c*} \tilde{\nu}^c_{b L}
 \crn && -
\left[\left(\sqrt{2}\varepsilon_{ab} v +   \fr{\sqrt{2}}{6 }\mu_\rho
\la^\prime_{ab}v'\right)\tilde{\nu}_{aL}\tilde{\nu}^c_{bL}\right.\crn
&&\left. -\fr{1}{2} u w \left(\fr{\la_a \la_b}{9} +  \fr{g^2}{2}
\fr{cos 2\beta}{s^2_\beta}\right)
\tilde{\nu}^*_{aL}\tilde{\nu}_{bL}^c + H.c.\right].\label{sf28}\eea
It is to be noticed that the last term in (\ref{sf28}) is the
mass-like (in the second order of fields) lepton-number violating
($\De L = \pm 2$). It is similar to the neutrino Majorona mass term;
and this is a special feature of the supersymmetric version.

For the sake of convenience, we will use the following notation
\bea && A_{ab} = \fr{g^2}{2} \de_{a b}\left( H_3
 + \fr{1}{\sqrt{3}}H_8 - \fr{2 t^2}{3}H_1 \right)
  + M^2_{ab} + \fr 1 4 \mu_{0a} \mu_{0b}\crn &&
  +\fr{1}{18}v^2(\la_a \la_b+4
\la^\prime_{ca}\la^\prime_{cb})+ \fr{1}{18}\la_a \la_b  w
^2,\\
 &&
 G_{ab} =  -g^2 \de_{a b}\left( \fr{1}{\sqrt{3}}H_8 +
 \fr{ t^2}{3}H_1 \right) + M^2_{ab} + \fr 1 4 \mu_{0a} \mu_{0b}
 \crn &&+\fr{1}{18}v^2(\la_a \la_b+4
\la^\prime_{ca}\la^\prime_{cb})+ \fr{1}{18}\la_a \la_b  u^2,\\
 &&
E_{ab} = - \sqrt{2}\left(\varepsilon_{ab} v +   \fr{1}{6 }\mu_\rho
\la^\prime_{ab}v'\right).\label{sf29}\eea
 In the base $(\tilde{\nu}_{a L}, \tilde{\nu}_{b R})=(\tilde{\nu}_{1 L},
 \tilde{\nu}_{2 L},\tilde{\nu}_{3 L},
 \tilde{\nu}_{1 R},\tilde{\nu}_{2 R}, \tilde{\nu}_{3 R})$, the mass
 matrix is given by
 \bea
\left(%
\begin{array}{cccccc}
  A_{ab} &  E_{ab} \\
  E_{ab} &  G_{ab}
\end{array}%
\right).
 \label{sf29} \eea
 Eigenstates and eigenmasses in this case are completely analogous
 to the charged sleptons with  replacements:
 $B_{33} \Rightarrow A_{33}$, $ C_{33} \Rightarrow  G_{33}$
and $ D_{33} \Rightarrow  E_{33} $.

As before, ignoring mixing among sneutrinos of two first
generations, we get eigenmasses and eigenstates given in
Table~\ref{eigenn}.
\begin{table}[h]
\caption{Masses and eigenstates of charged sleptons}
\begin{center}
\begin{tabular}{|c|c|c|c|c|}
\hline
 Eigenstate &$\tilde{\nu}_{1 L}$ & $\tilde{\nu}_{2 L}$ &
$\tilde{\nu}_{1 R}$ & $\tilde{\nu}_{2 R}$
\\
\hline
 $(\textrm{Mass})^2$ &$A_{11}$ & $A_{22}$ & $G_{11}$ & $G_{22}$   \\
 \hline
\end{tabular}
\label{eigenn}
\end{center}
\end{table}
and two  other sneutrinos are \bea \tilde{\nu}_{\tau L} &=&
s_{\theta_n}\tilde{\nu}_{3 R}
- c_{\theta_n}\tilde{\nu}_{3 L}, \label{sf27an}\\
\tilde{\nu}_{ \tau R} &=& c_{\theta_n}\tilde{\nu}_{3
R}+s_{\theta_n}\tilde{\nu}_{3 L}, \label{sf27bn}\eea with
respective masses \bea
m^2_{\tilde{\nu}_{\tau L}}&=&  \fr 1 2 (A_{33} + G_{33} - \De_n),\\
m^2_{\tilde{\nu}_{\tau R}}&=&  \fr 1 2 (A_{33} + G_{33} + \De_n),
\label{sf27cn}\eea where \bea \De_n &=& \sqrt{(G_{33}- A_{33})^2 +
4E_{33}^2},\label{sf35n}\\
t_{2\theta_n}&=&\fr{2E_{33}}{G_{33}-A_{33}}.\label{sf36n}\eea

 With the mentioned assumption, we have \bea
 m^2_{\tilde{\nu}_{1 L}} &=&  M^2_{11} + \fr 1 4
\mu_{01}^2 + \fr{g^2}{2} \left( H_3
 + \fr{1}{\sqrt{3}}H_8 - \fr{2 t^2}{3}H_1 \right)\crn&&
 +\fr{1}{18}v^2(\la_1^2+4
\la'^2_{c1})+ \fr{1}{18}\la_1^2 w
^2,\label{sf37n}\\
 m^2_{\tilde{\nu}_{2 L}}& =& M^2_{22} + \fr 1 4 \mu_{02}^2 +
\fr{g^2}{2} \left( H_3
 + \fr{1}{\sqrt{3}}H_8 - \fr{2 t^2}{3}H_1 \right)\crn&&
 +\fr{1}{18}v^2(\la_2^2+4
\la'^2_{c2})+ \fr{1}{18}\la_2^2 w ^2,
 \label{sf39n}\\
   m^2_{\tilde{\nu}_{1R}} &= & M^2_{11} + \fr 1 4 \mu_{01}^2
 -g^2 \left( \fr{1}{\sqrt{3}}H_8 +
 \fr{ t^2}{3}H_1 \right)\crn&&
 +\fr{1}{18}v^2(\la_1^2+4
\la'^2_{c1})+ \fr{1}{18}\la_1^2 u ^2,\label{sf40n}\\
m^2_{\tilde{\nu}_{2R}} &=& M^2_{22} + \fr 1 4 \mu_{02}^2  - g^2
\left( \fr{1}{\sqrt{3}}H_8 + \fr{ t^2}{3}H_1 \right)\crn&&
 +\fr{1}{18}v^2(\la_2^2+4
\la'^2_{c2})+ \fr{1}{18}\la_2^2 u ^2.\label{sf41n}
 \eea
 For the highest (tau)  sneutrinos:
 \bea  m^2_{\tilde{\nu}_{\tau L}} & = & \fr {1}
{2}\left[2 M_{33}^2+ \fr 1 2 \mu_{03}^2+\fr{1}{9}v^2(\la_3^2+4
\la'^3_{c3})+ \fr{1}{18}\la_3^2 (w ^2+u ^2)\right.\crn &&\left.+
\fr{g^2}{2} \left( H_3
 - \fr{1}{\sqrt{3}}H_8 - \fr{4 t^2}{3}H_1 \right)
-\De_n\right],\label{sf44n}\\
 m^2_{\tilde{\nu}_{\tau R}} & = & \fr {1}
{2}\left[2M_{33}^2+\fr {1}{2} \mu_{03}^{2} +\fr{1}{9}v^2(\la_3^2+4
\la'^3_{c3})+ \fr{1}{18}\la_3^2 (w ^2+u ^2)\right.\crn &&\left.+
\fr{g^2}{2} \left( H_3
 - \fr{1}{\sqrt{3}}H_8 - \fr{4 t^2}{3}H_1 \right)
+\De_n\right],\label{sf45n} \eea with \be \De_n = \sqrt{
\left[\fr{g^2}{2} \left( H_3
 + \sqrt{3}H_8  \right)+\fr {1}{18}\la_3^2(w^2-u ^2)\right]^2 +
8 \left(\varepsilon_{33} v + \fr 1 6 \mu_\rho
\la^\prime_{33}v^\prime\right)^2}.\label{sf46n}\ee

Comparing (\ref{sf37}), (\ref{sf39}) with (\ref{sf37n}),
(\ref{sf39n}), we see that, besides $R\!\!\!\!/$ -- coefficients,
without $D$-term contribution, there are degeneracies among
left-handed and right-handed sneutrinos, namely, $\tilde{\nu}_{1
L}$ among $ \tilde{\nu}_{1 R}$ and $\tilde{\nu}_{2 L}$ among $
\tilde{\nu}_{2 R}$.

\section{\label{squarkm}Squark masses}

Due to the fact that the exotic quarks in the model under
consideration have electric charges equal to that of the ordinary
ones, squark mass mixing matrices are expected to be larger than $6
\times 6$.

\subsection{Squark mass Lagrangian}

As before, the $D$-term contribution is diagonal, and let us
denote it by $D_Q$: \bea D_Q & \equiv &  g^2[(SQ_3 + SaQ_3) \times
H_3 + (SQ_8 + SaQ_8) \times  H_8 + (SQ_4 + SaQ_4) \times  H_4]\crn
&& + g^2 t^2 \times  H_1 (SQ_1 + SaQ_1)\crn
 &  = &
 g^2 \left\{ \tilde{u}^*_{ 1 L} \tilde{u}_{ 1 L} \left(\fr 1 2 H_3
+ \fr{1}{2 \sqrt{3}}H_8 + \fr{t^2}{3} H_1\right) - \fr{2 t^2}{3}H_1
\tilde{u}^{c*}_{ 1 L} \tilde{u}^c_{ 1 L} \right. \crn && +
\tilde{d}^*_{ 1 L} \tilde{d}_{ 1 L} \left(- \fr 1 2 H_3 + \fr{1}{2
\sqrt{3}}H_8 + \fr{t^2}{3} H_1\right) + \fr{ t^2}{3}H_1
\tilde{d}^{c*}_{ 1 L} \tilde{d}^c_{ 1 L} \crn &&    + \tilde{u}_{
 L}^{\prime
*}\tilde{u}_{  L}^{\prime } \left( - \fr{1}{ \sqrt{3}}H_8 +
\fr{t^2}{3} H_1\right) - \fr{2 t^2}{3}H_1 \tilde{u}^{\prime c*}_{ L}
\tilde{u}^{ \prime c}_{  L}\crn && +
 \tilde{u}^*_{ \al L} \tilde{u}_{ \al L} \left(\fr 1 2 H_3
- \fr{1}{2 \sqrt{3}}H_8 \right)- \fr{2 t^2}{3}H_1 \tilde{u}^{c*}_{
\al L} \tilde{u}^c_{ \al L}  \crn && - \tilde{d}^*_{ \al L}
\tilde{d}_{ \al L} \left( \fr 1 2 H_3 + \fr{1}{2 \sqrt{3}}H_8
\right)+ \fr{ t^2}{3}H_1 \tilde{d}^{c*}_{ \al L} \tilde{d}^c_{ \al
L} \crn && +\left. \fr{1}{ \sqrt{3}}H_8 \tilde{d}_{\al
 L}^{\prime
*}\tilde{d}_{\al  L}^{\prime } + \fr{ t^2}{3}H_1 \tilde{d}^{\prime
c*}_{\al L} \tilde{d}^{ \prime c}_{\al  L} + \fr 1 2
H_4\left(\tilde{u}^*_{1 L}\tilde{u}^\prime_L - \tilde{d}^*_{\al
L}\tilde{d}^\prime_{ \al L} + H.c.\right)\right\}.
 \label{sf601s}\eea

 Relevant mass terms for squarks arisen from
 the $F, D$ -terms and the soft terms are given by
 \bea \mathcal{L}_{squarks} & =& D_Q + m^2_{Q1L}
 \widetilde{Q}_{1L}^\dagger \widetilde{Q}_{1
L}+m^2_{Q\alpha \bet L}\widetilde{Q}_{\alpha L}^\dagger
\widetilde{Q}_{\bet L}\crn &&+  m^2_{u_{ij}}\widetilde{u}_{iL}^{c
*} \widetilde{u}_{jL}^c + m^2_{d_{ij}}\widetilde{d}_{iL}^{c
*} \widetilde{d}_{jL}^c + m^2_{u^\prime
}\widetilde{u^\prime}_{L}^{c *} \widetilde{u^\prime}_{L}^c+
m^2_{d^\prime \al \bet}\widetilde{d^\prime}_{\al L }^{c *}
\widetilde{d^\prime}_{\bet L}^c \crn &&
+\left[p_{i}\widetilde{Q}_{1L}\chi^\prime\widetilde{u}^c_{iL}
+p\widetilde{Q}_{1L}\chi^\prime\widetilde{ u^\prime}^c_{L} + p_{
\alpha i}\widetilde{Q}_{\alpha
L}\rho\widetilde{u}^c_{iL}+r_{\al}\widetilde{Q}_{\alpha
L}\rho\widetilde{u^\prime}^c_{L}+
h_{i}\widetilde{Q}_{1L}\rho^\prime\widetilde{d}^c_{iL} \right.\crn
&&+
h^\prime_{\al}\widetilde{Q}_{1L}\rho^\prime\widetilde{d^\prime}^c_{\al
L} +h_{\al i}\widetilde{Q}_{\alpha L}\chi\widetilde{d}^c_{iL}+
h^\prime_{\alpha\beta}\widetilde{Q}_{\alpha
L}\chi\widetilde{d^\prime}^c_{\beta L}\crn && +\fr 1 6 \mu_\chi
\chi^* (\kappa_{i} \tilde{Q}_{1L} \tilde{u}^{c}_{iL} + \kappa^\prime
\tilde{Q}_{1L} \tilde{u}^{\prime c}_L ) \crn && +\fr 1 6 \mu_\chi
\left( \Pi_{\alpha i} \chi^{\prime *}. \tilde{Q}_{\alpha L }
\tilde{d}^{c}_{iL} + \Pi^\prime_{\alpha \bet} \chi^{\prime *}.
\tilde{Q}_{\alpha L } \tilde{d}^{\prime c}_{\bet L}\right)\crn &&
+\fr 1 6 \mu_\rho \left(\pi_{ \alpha i}
 \rho^{\prime *}. \tilde{Q}_{\alpha L }\tilde{u}^{c}_{iL} +
\pi_{\alpha}^{\prime}  \rho^{\prime *}.\tilde{Q}_{\alpha L
}\tilde{u}^{\prime c}_{L} \right)\crn && +\fr 1 6 \mu_\rho \rho^*
\left( \vartheta_{i}\tilde{Q}_{1L} \tilde{d}^{c}_{iL} +
\vartheta^\prime_{ \alpha}\tilde{Q}_{1L} \tilde{d}^{\prime
c}_{\alpha L} \right) \crn &&+ \fr 1 6  \mu_{0a}
 \chi^{\prime \si *} (\xi_{a \alpha j}
 \tilde{Q}_{\alpha L \si} \tilde{d}^{c}_{jL}+
\xi^\prime_{a\alpha \beta} \tilde{Q}_{\alpha L \si}
\tilde{d}^{\prime c}_{\beta L}) \crn &&\left.+\fr{1}{9}\la_a
(\xi_{a\al i}\epsilon\tilde{Q}_{\al L}\chi^*\rho^*.
\tilde{d}^{c}_{iL}+\xi'_{a\al \bet}\epsilon\tilde{Q}_{\al
L}\chi^*\rho^*. \tilde{d}'^{c}_{\bet L}) +H.c.\right]\crn &&
+\fr{1}{18} \left[(w^{\prime 2} + u^{\prime 2})
(\kappa_{i}\kappa_{j} \tilde{u}^{c *}_{iL}\tilde{u}^{c}_{jL} +
\kappa^{\prime 2} \tilde{u}^{\prime c *}_L \tilde{u}^{\prime c}_L
+ \kappa_{i}\kappa^\prime\tilde{u}^{c *}_{iL}\tilde{u}^{\prime
c}_L + \kappa_{i}\kappa^\prime\tilde{u}^{c }_{iL}\tilde{u}^{\prime
c *}_L )\right. \crn && \left. + v^{\prime
2}(\vartheta_{i}\vartheta_{j} \tilde{d}^{c
*}_{iL}\tilde{d}^{c}_{jL} + \vartheta_{ \al}^{\prime }\vartheta_{
\bet}^{\prime }\tilde{d}^{\prime c *}_{\al L} \tilde{d}^{\prime
c}_{\bet L}  + \vartheta_{i} \vartheta_{ \al}^{\prime
}\tilde{d}^{c *}_{iL}\tilde{d}^{\prime c}_{\al L} + \vartheta_{i}
\vartheta_{ \al}^{\prime }\tilde{d}^{c }_{iL}\tilde{d}^{\prime c
*}_{\al L})
 \right]\crn && +
  \fr {1}{18}\left[v^2(\pi_{ \al i}\pi_{ \al j}
\tilde{u}^{c *}_{iL} \tilde{u}^{c }_{jL} + \pi_{\al}^{\prime
2}\tilde{u}^{\prime c *}_L
 \tilde{u}^{\prime c}_L + \pi_{ \al i}
 \pi_{ \al}^{\prime}\tilde{u}^{c *}_{iL}
\tilde{u}^{\prime c}_L + \pi_{ \al i}\pi_{
\al}^{\prime}\tilde{u}^{c }_{iL} \tilde{u}^{\prime c *}_L
 )\right.\crn && + (w^{ 2} + u^{ 2})(\Pi_{ \al i}\Pi_{ \al j}
 \tilde{d}^{ c *}_{iL}\tilde{d}^{ c}_{jL} + \Pi_{ \al \bet}^{\prime }
 \Pi_{ \al \de}^{\prime }\tilde{d}^{\prime c *}_{\bet L}
 \tilde{d}^{\prime c}_{\de L}\crn && \left.
 + \Pi_{ \al i}\Pi_{  \al \bet}^{\prime }
 \tilde{d}^{ c *}_{iL}\tilde{d}^{\prime c}_{\bet L}
 + \Pi_{ \al i}\Pi_{  \al \bet}^{\prime }
 \tilde{d}^{ c }_{iL}\tilde{d}^{\prime c *}_{\bet L})\right]\crn &&+ \fr
{1}{18} \left[u'^2\kappa_{i}^2
\widetilde{u}_{1L}^*\widetilde{u}_{1L}+  w'^2\kappa_{i}^2
\widetilde{u}'^*_{L}\widetilde{u}'_{L} +v^2 \pi_{ \al i} \pi_{
\bet i} \widetilde{u}_{\al L}\widetilde{u}^*_{\bet L}\right.\crn
&&\left.+(u' w' \kappa_{i}^2\widetilde{u}_{1L}^*\widetilde{u'}_{L}
-u' v \kappa_{i}\pi_{ \al i}\widetilde{u}_{1L}^*\widetilde{u}_{\al
L} -w'v \kappa_{i}\pi_{ \al
i}\widetilde{u}'^*_{L}\widetilde{u}_{\al L} +H.c. )\right] \crn &&
 +\fr {1}{18}
\left[u'^2\kappa'^2 \widetilde{u}_{1L}^*\widetilde{u}_{1L}+
w'^2\kappa'^2 \widetilde{u}'^*_{L}\widetilde{u}'_{L} +v^2 \pi'_{ \al
} \pi'_{ \bet } \widetilde{u}_{\al L}\widetilde{u}^*_{\bet
L}\right.\crn &&\left.+(u' w'
\kappa'^2\widetilde{u}_{1L}^*\widetilde{u'}_{L} -u' v \kappa'\pi'_{
\al}\widetilde{u}_{1L}^*\widetilde{u}_{\al L} -w'v \kappa'\pi'_{
\al}\widetilde{u}'^*_{L}\widetilde{u}_{\al L} +H.c. )\right]
\crn&&+\fr {1}{18} \left[v'^2 \vartheta_i^2
\widetilde{d}_{1L}\widetilde{d}_{1L}^* +u^2 \Pi_{\al i}\Pi_{\bet
i}\widetilde{d}_{\al L}\widetilde{d}^*_{\bet L} +w ^2 \Pi_{\al
i}\Pi_{\bet i}\widetilde{d}'_{\al L}\widetilde{d}'^*_{\bet
L}+\right.\crn &&\left. +(u w \Pi_{\al i}\Pi_{\bet
i}\widetilde{d}_{\al L}\widetilde{d}'^*_{\bet L}+v'u
\vartheta_i\Pi_{\al i}\widetilde{d}_{1 L}\widetilde{d}^*_{\al L}+v'w
\vartheta_i\Pi_{\al i}\widetilde{d}_{1 L}\widetilde{d}'^*_{\al L}
+H.c.)\right]\crn&&+\fr {1}{18} \left[v'^2 \vartheta'^2_\de
\widetilde{d}_{1L}\widetilde{d}_{1L}^* +u^2 \Pi'_{\al \de}\Pi'_{\bet
\de }\widetilde{d}_{\al L}\widetilde{d}^*_{\bet L} +w ^2 \Pi'_{\al
\de}\Pi'_{\bet \de}\widetilde{d}'_{\al L}\widetilde{d}'^*_{\bet
L}+\right.\crn &&\left. +(u w \Pi'_{\al \de}\Pi'_{\bet
\de}\widetilde{d}_{\al L}\widetilde{d}'^*_{\bet L}+v'u
\vartheta'_\de\Pi'_{\al \de}\widetilde{d}_{1 L}\widetilde{d}^*_{\al
L}+v'w \vartheta'_\de\Pi'_{\al \de}\widetilde{d}_{1
L}\widetilde{d}'^*_{\al L} +H.c.)\right]\crn&=&  D_Q +
m^2_{u_{ij}}\widetilde{u}_{iL}^{c *} \widetilde{u}_{jL}^c +
m^2_{d_{ij}}\widetilde{d}_{iL}^{c *} \widetilde{d}_{jL}^c +
m^2_{u^\prime }\widetilde{u^\prime}_{L}^{c *}
\widetilde{u^\prime}_{L}^c+ m^2_{d^\prime \al
\bet}\widetilde{d^\prime}_{\al L }^{c *} \widetilde{d^\prime}_{\bet
L}^c \crn && + \left[\fr{p_{i}}{\sqrt{2}} \left(u^\prime
\widetilde{u}_{1L}+w^\prime
\widetilde{u}_{L}^\prime\right)\widetilde{u}^c_{iL}
+\fr{p}{\sqrt{2}}\left(u^\prime \widetilde{u}_{1L}+w^\prime
\widetilde{u}_{L}^\prime\right)\widetilde{ u^\prime}^c_{L}\right.
\crn && - \fr{vp_{ \alpha i}}{\sqrt{2}}\widetilde{u}_{\alpha
L}\widetilde{u}^c_{iL}- \fr{vr_{\al}}{\sqrt{2}}\widetilde{u}_{\alpha
L}\widetilde{u^\prime}^c_{L}+
\fr{v'h_{i}}{\sqrt{2}}\widetilde{d}_{1L}\widetilde{d}^c_{iL} \crn
&&+ \fr{v'h^\prime_{\al}}{\sqrt{2}}
\widetilde{d}_{1L}\widetilde{d^\prime}^c_{\al L} +\fr{h_{\al
i}}{\sqrt{2}}(u\widetilde{d}_{\alpha L}+w\widetilde{d}'_{\alpha
L})\widetilde{d}^c_{iL}+
\fr{h^\prime_{\alpha\beta}}{\sqrt{2}}\left(u\widetilde{d}_{\alpha
L}+w\widetilde{d}'_{\alpha L}\right)\widetilde{d^\prime}^c_{\beta
L}\crn && +\fr {1}{ 6\sqrt{2}} \mu_\chi  (\kappa_{i}
(u\tilde{u}_{1L}+w\tilde{u}'_{L}) \tilde{u}^{c}_{iL} + \kappa^\prime
(u\tilde{u}_{1L}+w\tilde{u}'_{L})\tilde{u}^{\prime c}_L ) \crn &&
+\fr {1}{ 6\sqrt{2}} \mu_\chi \left( \Pi_{\alpha i} (
u'\widetilde{d}_{\alpha L}+w'\widetilde{d}'_{\alpha L})
\tilde{d}^{c}_{iL} + \Pi^\prime_{\alpha \bet}(
u'\widetilde{d}_{\alpha L}+w'\widetilde{d}'_{\alpha L})
\tilde{d}^{\prime c}_{\bet L}\right)\crn && -\fr {1}{ 6\sqrt{2}}
\mu_\rho \left(\pi_{ \alpha i}
 v'\tilde{u}_{\alpha L }\tilde{u}^{c}_{iL} +
\pi_{\alpha}^{\prime}  v'\tilde{u}_{\alpha L }\tilde{u}^{\prime
c}_{L} \right)+\fr {v}{ 6\sqrt{2}} \mu_\rho \left(
\vartheta_{i}\tilde{d}_{1L} \tilde{d}^{c}_{iL} + \vartheta^\prime_{
\alpha}\tilde{d}_{1L} \tilde{d}^{\prime c}_{\alpha L} \right) \crn
&& + \fr {1} {6\sqrt{2}}  \mu_{0a}
 (\xi_{a \alpha j} (u'\tilde{d}_{\alpha L}
 +w'\tilde{d}'_{\alpha L}) \tilde{d}^{c}_{jL}+
\xi^\prime_{a\alpha \beta} (u'\tilde{d}_{\alpha
L}+w'\tilde{d}'_{\alpha L}) \tilde{d}^{\prime c}_{\beta L})
 \crn
&&\ +\fr{1}{18}\la_a w v (\xi_{a\al i}\tilde{d}_{\alpha
L}\tilde{d}^c_{i L}+\xi'_{a\al \bet}\tilde{d}_{\alpha
L}\tilde{d}'^c_{\bet L})\crn &&+\fr{1}{18}\la_a u v (\xi_{a\al
i}\tilde{d}'_{\alpha L}\tilde{d}^c_{i L}+\xi'_{a\al
\bet}\tilde{d}'_{\alpha L}\tilde{d}'^c_{\bet L})\crn &&+\fr
{1}{18}u' w'
(\kappa_{i}^2+\kappa'^2)\widetilde{u}_{1L}^*\widetilde{u'}_{L}
-\fr {1}{18}u' v (\kappa_{i}\pi_{ \al i}+\kappa'\pi'_{ \al
})\widetilde{u}_{1L}^*\widetilde{u}_{\al L}\crn && - \fr
{1}{18}w'v (\kappa_{i}\pi_{ \al i}+\kappa'\pi'_{ \al
})\widetilde{u}'^*_{L}\widetilde{u}_{\al L} +\fr {1}{18}u w
(\Pi_{\al i}\Pi_{\bet i}+\Pi'_{\al \de}\Pi'_{\bet
\de})\widetilde{d}_{\al L}\widetilde{d}'^*_{\bet L}\crn
&&\left.+\fr {1}{18} v'u (\vartheta_i\Pi_{\al i}+
\vartheta'\Pi'_{\al \de})\widetilde{d}_{1 L}\widetilde{d}^*_{\al
L}+\fr {1}{18}v'w (\vartheta_i\Pi_{\al i}+ \vartheta'\Pi'_{\al
\de})\widetilde{d}_{1 L}\widetilde{d}'^*_{\al L} +H.c. \right]\crn
&& +\fr{1}{18}\left[(w^{\prime 2} + u^{\prime 2})
(\kappa_{i}\kappa_{j} \tilde{u}^{c *}_{iL}\tilde{u}^{c}_{jL} +
\kappa^{\prime 2} \tilde{u}^{\prime c *}_L \tilde{u}^{\prime c}_L
+ \kappa_{i}\kappa^\prime\tilde{u}^{c *}_{iL}\tilde{u}^{\prime
c}_L + \kappa_{i}\kappa^\prime\tilde{u}^{c }_{iL}\tilde{u}^{\prime
c *}_L )\right. \crn && \left. + v^{\prime
2}(\vartheta_{i}\vartheta_{j} \tilde{d}^{c
*}_{iL}\tilde{d}^{c}_{jL} + \vartheta_{ \al}^{\prime }\vartheta_{
\bet}^{\prime }\tilde{d}^{\prime c *}_{\al L} \tilde{d}^{\prime
c}_{\bet L}  + \vartheta_{i} \vartheta_{ \al}^{\prime
}\tilde{d}^{c *}_{iL}\tilde{d}^{\prime c}_{\al L} + \vartheta_{i}
\vartheta_{ \al}^{\prime }\tilde{d}^{c }_{iL}\tilde{d}^{\prime c
*}_{\al L})
 \right]\crn && +
  \fr {1}{18}\left[v^2(\pi_{ \al i}\pi_{ \al j}
\tilde{u}^{c *}_{iL} \tilde{u}^{c }_{jL} + \pi_{\al}^{\prime
2}\tilde{u}^{\prime c *}_L
 \tilde{u}^{\prime c}_L + \pi_{ \al i}\pi_{ \al}^{\prime}
 \tilde{u}^{c *}_{iL}
\tilde{u}^{\prime c}_L + \pi_{ \al i}\pi_{
\al}^{\prime}\tilde{u}^{c }_{iL} \tilde{u}^{\prime c *}_L
 )\right. \crn && + (w^{ 2} + u^{ 2})(\Pi_{ \al i}\Pi_{ \al j}
 \tilde{d}^{ c *}_{iL}\tilde{d}^{ c}_{jL} + \Pi_{ \al \bet}^{\prime }
 \Pi_{ \al \de}^{\prime }\tilde{d}^{\prime c *}_{\bet L}
 \tilde{d}^{\prime c}_{\de L}\crn &&
 \left. + \Pi_{ \al i}\Pi_{  \al \bet}^{\prime }
 \tilde{d}^{ c *}_{iL}\tilde{d}^{\prime c}_{\bet L}
 + \Pi_{ \al i}\Pi_{  \al \bet}^{\prime }
 \tilde{d}^{ c }_{iL}\tilde{d}^{\prime c *}_{\bet L}
 )\right]\crn &&
 +m^2_{Q1L}(\tilde{u}^{ *}_{1L}\tilde{u}_{1L} + \tilde{d}^{
*}_{1L}\tilde{d}_{1L} + \tilde{u}^{\prime
*}_{L}\tilde{u}^\prime_{L}) \crn && +m^2_{Q\alpha \bet L}
(\tilde{u}^{ *}_{\al L}\tilde{u}_{\bet L} + \tilde{d}^{ *}_{\al
L}\tilde{d}_{\bet L} + \tilde{d}^{\prime *}_{\al
L}\tilde{d}^\prime_{\bet L} ) \crn &&+ \fr {1}{18}
\left[u'^2(\kappa_{i}^2+\kappa'^2)
\widetilde{u}_{1L}^*\widetilde{u}_{1L}+ w'^2(\kappa_{i}^2+\kappa'^2)
\widetilde{u}'^*_{L}\widetilde{u}'_{L} +v^2 (\pi_{ \al i} \pi_{ \bet
i}+\pi'_{ \al } \pi'_{ \bet }) \widetilde{u}_{\al
L}\widetilde{u}^*_{\bet L}\right] \crn &&+\fr {1}{18} \left[v'^2
(\vartheta_i^2+\vartheta'^2_\de)
\widetilde{d}_{1L}\widetilde{d}_{1L}^* +u^2 (\Pi_{\al i}\Pi_{\bet
i}+\Pi'_{\al \de}\Pi'_{\bet \de})\widetilde{d}_{\al
L}\widetilde{d}^*_{\bet L} \right.\crn &&\left.+w ^2 (\Pi_{\al
i}\Pi_{\bet i}+\Pi'_{\al \de}\Pi'_{\bet \de})\widetilde{d}'_{\al
L}\widetilde{d}'^*_{\bet L}\right]\crn& = & D_Q +
\left[m^2_{u_{ij}}+\fr {1}{18}(w^{\prime 2} + u^{\prime 2})
\kappa_{i}\kappa_{j}+\fr {1}{18} v^2\pi_{ \al i}\pi_{ \al
j}\right]\widetilde{u}_{iL}^{c *} \widetilde{u}_{jL}^c \crn&&+
\left[m^2_{d_{ij}}+\fr {1}{18}  v^{\prime
2}\vartheta_{i}\vartheta_{j} +\fr {1}{18}(w^{ 2} + u^{ 2})\Pi_{ \al
i}\Pi_{ \al j} \right]\widetilde{d}_{iL}^{c *}
\widetilde{d}_{jL}^c\crn&& +\left[m^2_{u^\prime }+ \fr {1}{18}
\kappa^{\prime 2}(u'^2+w'^2)+\fr {1}{18}v^2\pi_{\al}^{\prime
2}\right]\widetilde{u^\prime}_{L}^{c *}
\widetilde{u^\prime}_{L}^c\crn&& + \left[m^2_{d^\prime \al \bet}+
\fr {1}{18} v^{\prime 2} \vartheta_{ \al}^{\prime }\vartheta_{
\bet}^{\prime }+\fr {1}{18} (w^{ 2} + u^{ 2}) \Pi_{ \de
\bet}^{\prime }
 \Pi_{\de \al }^{\prime }\right]\widetilde{d^\prime}_{\al
L }^{c *} \widetilde{d^\prime}_{\bet L}^c \crn && +[m^2_{Q1L}+\fr
{1}{18}u'^2(\kappa_{i}^2+\kappa'^2)]\tilde{u}^{
*}_{1L}\tilde{u}_{1L} \crn &&+[m^2_{Q1L}+\fr {1}{18} v'^2
(\vartheta_i^2+\vartheta'^2_\de)] \tilde{d}^{
*}_{1L}\tilde{d}_{1L} \crn &&+ [m^2_{Q1L}+\fr
{1}{18}w'^2(\kappa_{i}^2+\kappa'^2)]\tilde{u}^{\prime
*}_{L}\tilde{u}^\prime_{L} \crn && +[m^2_{Q\alpha \bet
L}+\fr{1}{18}v^2 (\pi_{ \al i} \pi_{ \bet i}+\pi'_{ \al } \pi'_{
\bet })]\tilde{u}^{ *}_{\al L}\tilde{u}_{\bet L} \crn &&
+[m^2_{Q\alpha \bet L}+\fr{1}{18}u^2 (\Pi_{\al i}\Pi_{\bet
i}+\Pi'_{\al \de}\Pi'_{\bet \de})] \tilde{d}^{ *}_{\al
L}\tilde{d}_{\bet L} \crn && +[m^2_{Q\alpha \bet L}+\fr{1}{18}w^2
(\Pi_{\al i}\Pi_{\bet i}+\Pi'_{\al \de}\Pi'_{\bet \de}) ]
\tilde{d}^{\prime *}_{\al L}\tilde{d}^\prime_{\bet L} \crn &&
+\left\{\left(\fr{u^\prime p_{i}}{\sqrt{2}}+\fr { u\mu_\chi
\kappa_{i}}{
6\sqrt{2}}\right)\widetilde{u}_{1L}\widetilde{u}^c_{iL}
+\left(\fr{w^\prime p_{i}}{\sqrt{2}}+\fr { w \mu_\chi \kappa_{i}}{
6\sqrt{2}}\right)\widetilde{u}_{L}^\prime\widetilde{u}^c_{iL}
+\left(\fr{w^\prime p}{\sqrt{2}}+\fr { w\mu_\chi \kappa^\prime}{
6\sqrt{2}}\right)\widetilde{u}_{L}^\prime\widetilde{u}^{\prime
c}_{L}\right. \crn&& + \left(\fr{u^\prime p}{\sqrt{2}}+\fr {
u\mu_\chi \kappa^\prime}{
6\sqrt{2}}\right)\widetilde{u}_{1L}\widetilde{u}^{\prime c}_{L}-
\left(\fr{vp_{ \alpha i}}{\sqrt{2}}+ \fr {v^\prime\mu_\rho \pi_{
\alpha i}}{ 6\sqrt{2}} \right)\widetilde{u}_{\alpha
L}\widetilde{u}^c_{iL}- \left(\fr{vr_{ \alpha }}{\sqrt{2}}+ \fr
{v^\prime\mu_\rho \pi'_{ \alpha }}{ 6\sqrt{2}}
\right)\widetilde{u}_{\alpha L}\widetilde{u}'^c_{L}\crn && +
\left(\fr{v'h_{i}}{\sqrt{2}}+\fr {v\mu_\rho \vartheta_{  i}}{
6\sqrt{2}} \right)\widetilde{d}_{1L}\widetilde{d}^c_{iL}
+\left(\fr{v'h'_{\alpha}}{\sqrt{2}}+\fr {v\mu_\rho \vartheta'_{
\alpha}}{ 6\sqrt{2}}
\right)\widetilde{d}_{1L}\widetilde{d}'^c_{\alpha L}\crn &&+
\left(\fr{uh_{\al i}}{\sqrt{2}}+\fr {u'\mu_\chi \Pi_{\alpha i}}{
6\sqrt{2}} +\fr {u'\mu_{0a} \xi_{a\alpha i}}{
6\sqrt{2}}+\fr{1}{18}\la_a w v \xi_{a\al
i}\right)\widetilde{d}_{\alpha
L}\widetilde{d}^c_{iL}\crn&&+\left(\fr{wh_{\al i}}{\sqrt{2}}+\fr
{w'\mu_\chi \Pi_{\alpha i}}{ 6\sqrt{2}} +\fr {w'\mu_{0a}
\xi_{a\alpha i}}{ 6\sqrt{2}}+\fr{1}{18}\la_a u v \xi_{a\al
i}\right)\widetilde{d}'_{\alpha L}\widetilde{d}^c_{iL}\crn&&+
\left(\fr{uh^\prime_{\alpha\beta}}{\sqrt{2}}+\fr {u'\mu_\chi
\Pi'_{\alpha \beta}}{ 6\sqrt{2}}+\fr{u'\mu_{0a}\xi^\prime_{a\alpha
\beta}}{ 6\sqrt{2}}+\fr{1}{18}\la_a w v \xi'_{a\al \bet}
\right)\widetilde{d}_{\alpha L}\widetilde{d}'^c_{\beta L}\crn &&
+\left(\fr{wh^\prime_{\alpha\beta}}{\sqrt{2}}+\fr {w'\mu_\chi
\Pi'_{\alpha \beta}}{ 6\sqrt{2}}+\fr{w'\mu_{0a}\xi^\prime_{a\alpha
\beta}}{ 6\sqrt{2}}+\fr{1}{18}\la_a u v \xi'_{a\al \bet}
\right)\widetilde{d}'_{\alpha L}\widetilde{d}'^c_{\beta L}\crn &&
+ \fr {1}{18}\left[(w'^{ 2} + u'^{ 2})
\kappa_{i}\kappa^\prime+v^2\pi_{ \al i}\pi_{
\al}^{\prime}\right]\tilde{u}^{c }_{iL} \tilde{u}^{\prime c
*}_L\crn&&+ \fr {1}{18}\left[ v^{\prime 2} \vartheta_{i}
\vartheta_{ \bet}^{\prime }+(w^{ 2} + u^{ 2})\Pi_{ \al i}\Pi_{ \al
\bet}^{\prime }\right]\tilde{d}^{ c }_{iL}\tilde{d}^{\prime c
*}_{\bet L}  \crn &&+ \fr 1 2
\widetilde{u}_{1L}^*\widetilde{u'}_{L}\left[\fr {1}{9}u' w'
(\kappa_{i}^2+\kappa'^2) - \fr{g^2}{2} u w \fr{\cos
2\beta}{s^2_\beta}\right]
 -\fr {1}{18}u' v (\kappa_{i}\pi_{ \al
i}+\kappa'\pi'_{ \al })\widetilde{u}_{1L}^*\widetilde{u}_{\al L}\crn
&& - \fr {1}{18}w'v (\kappa_{i}\pi_{ \al i}+\kappa'\pi'_{ \al
})\widetilde{u}'^*_{L}\widetilde{u}_{\al L} +\fr {1}{18}u w
(\Pi_{\al i}\Pi_{\bet i}+\Pi'_{\al \de}\Pi'_{\bet
\de})\widetilde{d}_{\al L}\widetilde{d}'^*_{\bet L}\crn &&+\fr
{1}{18} v'u (\vartheta_i\Pi_{\al i}+ \vartheta'\Pi'_{\al
\de})\widetilde{d}_{1 L}\widetilde{d}^*_{\al L}+\fr {1}{18}v'w
(\vartheta_i\Pi_{\al i}+ \vartheta'\Pi'_{\al \de})\widetilde{d}_{1
L}\widetilde{d}'^*_{\al L}\crn && \left. + \fr{g^2}{4} u w \fr{\cos
2\beta}{s^2_\beta} \widetilde{d}_{\al L}\widetilde{d}'^*_{\al L} +
H.c. \right\}\label{sqmassterm} \eea Looking at (\ref{sqmassterm})
we see that there is mixing among ordinary squarks with exotic
squarks (with primes) and this produces $8 \times 8$ matrix for
up-squarks and $10 \times10$ matrix for down-squaks. Noting that
exotic squarks carry lepton number $\pm 2$, we conclude that
coefficients of the mixture among ordinary and exotic squarks are
very small. In the following, we will neglect such mixing.

\subsection{The lepton number conservation limit}

Remind that, in the SM, neutrinos are massless and lepton number is
conserved. Let us consider the SM limit. The lepton-number
conservation conditions imposed to (\ref{sqmassterm}) are the
following: \bea && w^\prime p_{i} +\fr { w \mu_\chi \kappa_{i}}{ 6}
= u^\prime p +\fr { u\mu_\chi \kappa^\prime}{ 6} = vr_{ \alpha }+
\fr {v^\prime\mu_\rho \pi'_{ \alpha }}{ 6}  = v'h'_{\alpha}+\fr
{v\mu_\rho \vartheta'_{
\alpha}}{ 6}  = 0,\label{thuy912}\\
&& wh_{\al i}+\fr {w'\mu_\chi \Pi_{\alpha i}}{ 6} +\fr {w'\mu_{0a}
\xi_{a\alpha i}}{ 6} +\fr{1}{18}\la_a u v \xi_{a\al
i}= 0,\label{th2}\\
&& uh^\prime_{\alpha\beta}+\fr {u'\mu_\chi \Pi'_{\alpha \beta}}{
6}+\fr{u'\mu_{0a}\xi^\prime_{a\alpha \beta}}{ 6}
+\fr{1}{18}\la_a w v \xi'_{a\al \bet} = 0,\label{th3}\\
&&(w'^{ 2} + u'^{ 2}) \kappa_{i}\kappa^\prime+v^2\pi_{ \al i}\pi_{
\al}^{\prime}= v^{\prime 2} \vartheta_{i} \vartheta_{ \bet}^{\prime
}+(w^{ 2} + u^{ 2})\Pi_{ \al i}\Pi_{  \al \bet}^{\prime }=0,\\
&& \fr {1}{9}u' w' (\kappa_{i}^2+\kappa'^2) - \fr {g^2}{ 2}  u w
\fr{\cos 2\beta}{s^2_\beta} = w' v(\kappa_{i}\pi_{ \al
i}+\kappa'\pi'_{ \al })  =0,
 \\&&\fr 1 2 u w \left[\fr 1 9 (\Pi_{\al i}\Pi_{\al
i}+\Pi'_{\al \de}\Pi'_{\al \de}) +\fr {g^2}{ 2} \fr{\cos
2\beta}{s^2_\beta}\right]= v' w(\vartheta_i\Pi_{\al i}+
\vartheta'\Pi'_{\al \de})=0,
\\&& u w(\Pi_{2 i}\Pi_{3 i}+\Pi'_{2 \de}\Pi'_{3 \de})
=0.\label{sqrparity} \eea With imposition lepton-number
conservation, the  difficulties of large squark mixing,  will be
very much eased. Let us denote
 \bea \overline{A}_{ij}&=& m^2_{u_{ij}}+\fr
{1}{18}(w^{\prime 2} + u^{\prime 2}) \kappa_{i}\kappa_{j}+\fr
{1}{18} v^2\pi_{ \al i}\pi_{ \al j}- \fr 2 3 g^2 t^2 H_1\de_{ij},\\
\overline{B}_{ij}&=&m^2_{d_{ij}}+\fr {1}{18}  v^{\prime
2}\vartheta_{i}\vartheta_{j} +\fr {1}{18}(w^{ 2} + u^{ 2})\Pi_{ \al
i}\Pi_{ \al j} + \fr 1 3 g^2 t^2 H_1\de_{ij},\\
 \overline{C}&=& m^2_{u^\prime }+
 \fr {1}{18} \kappa^{\prime 2}(u'^2+w'^2)+\fr
{1}{18}v^2\pi_{\al}^{\prime 2}- \fr 2 3 g^2 t^2 H_1,\\
\overline{P}_{\al \bet}&=& m^2_{d^\prime \al \bet}+ \fr {1}{18}
v^{\prime 2} \vartheta_{ \al}^{\prime }\vartheta_{ \bet}^{\prime
}+\fr {1}{18} (w^{ 2} + u^{ 2}) \Pi_{ \de \bet}^{\prime }
 \Pi_{\de \al }^{\prime } + \fr 1 3 g^2 t^2 H_1\de_{\al \bet},\\
 \overline{E}_{u_{1L}}&=&m^2_{Q1L}+\fr {1}{18}  u^{\prime
2}(\kappa_{i}^2+\kappa^{\prime 2}) + g^2 \left(\fr{1}{2}H_3
+ \fr{1}{2\sqrt{3}}H_8 +  \fr 1 3  t^2 H_1\right),\label{cte}\\
\overline{E}_{d_{1L}}&=&m^2_{Q1L}+\fr {1}{18} v^{\prime 2}(
\vartheta^2_i+\vartheta^{\prime 2}_\alpha) + g^2 \left(-\fr{1}{2}H_3
+ \fr{1}{2\sqrt{3}}H_8 +  \fr 1 3  t^2 H_1\right),\\
\overline{E}_{u^\prime_{L}}&=&m^2_{Q1L}+\fr {1}{18} w^{\prime
2}(\kappa_{i}^2+\kappa^{\prime 2}) - g^2 \left(
 \fr{1}{\sqrt{3}}H_8 -  \fr 1 3  t^2 H_1\right),\\
 \overline{F}_{u_{\al \bet}}&=&m^2_{Q\alpha \bet L}+
 \fr {1}{18}v^2( \pi_{ \al i}\pi_{ \bet i}+
\pi^\prime_{ \al }\pi^\prime_{ \bet })+ g^2 \de_{\al \bet}
\left(\fr{1}{2}H_3 -
\fr{1}{2\sqrt{3}}H_8 \right), \\
\overline{F}_{d_{\al \bet}}&=&m^2_{Q\alpha \bet L}+
 \fr {1}{18} u^ 2(\Pi_{\beta \de }^\prime\Pi_{ \alpha
 \de }^\prime+\Pi_{ \al i}\Pi_{
\bet i })  - g^2 \de_{\al \bet} \left(\fr{1}{2}H_3 +
\fr{1}{2\sqrt{3}}H_8 \right),\label{ctf} \\
\overline{F}_{d^\prime_{\al \bet}}&=&m^2_{Q\alpha \bet L}+ \fr
{1}{18} w^ 2(\Pi_{\beta \de }^\prime\Pi_{ \alpha\de }^\prime+\Pi_{
\al i}\Pi_{ \bet i })  + g^2 \de_{\al \bet}
\fr{1}{\sqrt{3}}H_8 \label{ctft} , \\
 \overline{G}_{\alpha i}&=&- \left(\fr{vp_{ \alpha i}}{\sqrt{2}}
 + \fr {v^\prime\mu_\rho
\pi_{ \alpha i}}{ 6\sqrt{2}} \right),\\
\overline{H}_{i}&=&\fr{v'h_{i}}{\sqrt{2}}+\fr {v\mu_\rho \vartheta_{
i}}{ 6\sqrt{2}},\\
 \overline{K}_{\al i}&=&\fr{uh_{\al
i}}{\sqrt{2}}+\fr {u'\mu_\chi \Pi_{\alpha i}}{ 6\sqrt{2}} +\fr
{u'\mu_{0a} \xi_{a\alpha i}}{ 6\sqrt{2}}+\fr{1}{18}\la_a w v
\xi_{a\al i},\\
\overline{N}_{\al \bet}&=&\fr{wh^\prime_{\alpha\beta}}{\sqrt{2}}+\fr
{w'\mu_\chi \Pi'_{\alpha \beta}}{
6\sqrt{2}}+\fr{w'\mu_{0a}\xi^\prime_{a\alpha \beta}}{
6\sqrt{2}}+\fr{1}{18}\la_a u v \xi'_{a\al \bet}\label{sf90}\eea
 Then the mass Lagrangian (\ref{sqmassterm})
 can be rewritten in the form
\bea \mathcal{L}_{squarks} & =& \overline{C}\tilde{u}'^{c
*}_{L}\tilde{u}^{\prime c}_L+
\overline{E}_{u^\prime_{L}}\tilde{u}'^{
*}_{L}\tilde{u}^{\prime}_L+
 \overline{E}_{u_{1L}}\tilde{u}^{
*}_{1L}\tilde{u}_{1L}+\overline{E}_{d_{1L}}\tilde{d}^{
*}_{1L}\tilde{d}_{1L}
\crn && +\overline{A}_{ij}\tilde{u}^{ c*}_{iL}\tilde{u}^c_{jL}
+\overline{P}_{\alpha\beta}\tilde{d}'^{ c*}_{\alpha
L}\tilde{d}^{\prime c}_{\beta L} +\overline{F}_{u_{\al \bet}}
\tilde{u}^{
*}_{\al L}\tilde{u}_{\bet L}\crn&& +\overline{B}_{ij}\tilde{d}^{
c*}_{iL}\tilde{d}^c_{jL} +\overline{F}_{d^\prime_{\al
\bet}}\tilde{d}'^{
*}_{\alpha L}\tilde{d}^{\prime }_{\beta L}
+\overline{F}_{d_{\al \bet}} \tilde{d}^{
*}_{\al L}\tilde{d}_{\bet
L}\crn&&+[\overline{K}_{\al i}\tilde{d}_{\al
L}\tilde{d}^c_{iL}+\overline{N}_{\al \bet}\tilde{d}'_{\al
L}\tilde{d}'^c_{\bet L}+\overline{G}_{\al i}\tilde{u}_{\al
L}\tilde{u}^c_{iL} +\overline{H}_{i}\tilde{d}_{1L}\tilde{d}^c_{iL}
+H.c.]\label{sf91}\eea From (\ref{sf91}) we see that all the exotic
squarks are decoupled of ordinary squarks. The reason of this is
that in the 3-3-1 models, the exotic quarks carry also lepton
number, while the ordinary ones do not, so their superpartners have
the same property.

Looking at (\ref{sf91}), we conclude that the $\tilde{u}'^c_{L}$ and
$\tilde{u}'_{L}$ gain masses respectively, \be
m^2_{\tilde{u}'^c_{L}} = \overline{C}, \hs
 m^2_{\tilde{u}'_{L}} =
 \overline{E}_{u^\prime_{L}}\label{sf92}\ee

 For the ordinary up-squarks, the $\tilde{u}_{1L}$ does not mix and
 gains mass:
 \be
 m^2_{\tilde{u}_{1L}} = \overline{E}_{u_{1L}}.
 \label{sf95}\ee
 The remaining up-squarks are all mixing and
 in the base $(\tilde{u}^*_{2L},\tilde{u}^*_{3L},\tilde{u}^c_{1L},
 \tilde{u}^c_{2L},\tilde{u}^c_{3L})$, the mass matrix is given by
 \bea\left(
       \begin{array}{ccccc}
         \overline{F}_{u_{22}} & \overline{F}_{u_{23}} & \overline{G}_{21}
          & \overline{G}_{22} & \overline{G}_{23} \\
          \overline{F}_{u_{32}} &\overline{F}_{u_{33}}
         &\overline{G}_{31} & \overline{G}_{32} & \overline{G}_{33} \\
          \overline{G}_{21} & \overline{G}_{31} & \overline{A}_{11}
         & \overline{A}_{21} & \overline{A}_{31} \\
          \overline{G}_{22} & \overline{G}_{32}
         & \overline{A}_{12} & \overline{A}_{22} & \overline{A}_{32} \\
          \overline{G}_{23} & \overline{G}_{33}
         & \overline{A}_{13} & \overline{A}_{23} & \overline{A}_{33}
       \end{array}
     \right).
\label{sf96}
 \eea
 For superpartners of the down-quarks ($q= - \fr 1 3$), we have: The
$\tilde{d}'^*_{2L}$, $\tilde{d}'^*_{3L}$, $\tilde{d}'^c_{2L}$ and
$\tilde{d}'^c_{3L}$ mix with mass matrix \bea \left(
  \begin{array}{cccc}
     \overline{F}_{d^\prime_{2 2}} & \overline{F}_{d^\prime_{2 3}}
     & \overline{N}_{22} &  \overline{N}_{23}\\
    \overline{F}_{d^\prime_{3 2}} & \overline{F}_{d^\prime_{33}}
     & \overline{N}_{32} &  \overline{N}_{33} \\
    \overline{N}_{22} &  \overline{N}_{32}
    & \overline{P}_{22} & \overline{P}_{32} \\
    \overline{N}_{23} &  \overline{N}_{33}
    & \overline{P}_{23} & \overline{P}_{33} \\
  \end{array}
\right).\label{sf93s}\eea
  For the ordinary down-squark,in the base
$(\tilde{d}^*_{1L},\tilde{d}^*_{2L},\tilde{d}^*_{3L},\tilde{d}^c_{1L},
 \tilde{d}^c_{2L},\tilde{d}^c_{3L})$, the mass matrix is defined
by\bea
\left(
  \begin{array}{cccccc}
    \overline{E}_{d_{1L}} & 0 & 0 & \overline{H}_1 &
     \overline{H}_2  & \overline{H}_3 \\
    0 & \overline{F}_{d_{22}} & \overline{F}_{d_{23}}
     & \overline{K}_{21} & \overline{K}_{22} & \overline{K}_{23} \\
    0 & \overline{F}_{d_{32}} & \overline{F}_{d_{33}}
    & \overline{K}_{31} & \overline{K}_{32} & \overline{K}_{33} \\
    \overline{H}_1 & \overline{K}_{21} & \overline{K}_{31}
    & \overline{B}_{11} & \overline{B}_{21} & \overline{B}_{31} \\
    \overline{H}_2& \overline{K}_{22} & \overline{K}_{32}
     & \overline{B}_{12} & \overline{B}_{22} & \overline{B}_{32} \\
    \overline{H}_3 & \overline{K}_{23} & \overline{K}_{33}
    & \overline{B}_{13}
    & \overline{B}_{23} & \overline{B}_{33} \\
  \end{array}
\right). \label{sf98s}\eea As in the MSSM, the ordinary
down-squarks mix by $6 \times 6$ matrix.

 It is to be noted that the decoupling of $\tilde{u}_{1L}$ is a
result of the condition for  minimum of Higgs potential: $u' / u =
w' / w$ and the absence  of $Q_{1 L}$ in two last  terms of Eq.
(\ref{w3}).

 In general, we
cannot deal with $5 \times 5$ and $6 \times 6$ matrices. Following
Refs \cite{martin} and \cite{pdg}, we expect the $\tilde{q}_L -
\tilde{q}_R $ mixing to be small, with possible exception of the
third-generation, where mixing can be enhanced by factors of $m_t$
and $m_b$. Keeping in mind this assumption, from Eq. (\ref{sf96})
to Eq. (\ref{sf98s}), we conclude that, all non-vanishing matrix
elements are: $ \overline{F}_{u_{22}}$, $ \overline{F}_{u_{33}}$,
$ \overline{A}_{11}$, $ \overline{A}_{22}$, $ \overline{A}_{33}$,
$ \overline{G}_{33}$,  $\overline{F}_{d'_{22}}$,
$\overline{F}_{d'_{33}}$,$ \overline{P}_{22}$, $
\overline{P}_{33}$,$\overline{N}_{33}$, $
\overline{E}_{d_{1L}}$,$\overline{F}_{d_{22}}$,
$\overline{F}_{d_{33}}$, $\overline{B}_{11}$, $
\overline{B}_{22}$, $ \overline{B}_{33}$ and $\overline{K}_{33}$
With the help of the above assumption, diagonalization of the mass
mixing matrices is quite easy. Our results are as follows: \ben
\item {\it For up-squarks}:

The  eigenmasses and eigenstates are given in
Table~\ref{eigensquarks}.
 \begin{table}[h]
\caption{Masses and eigenstates of  up-squarks}
\begin{center}
 \begin{tabular}{|c|c|c|c|c|c|c|}
   \hline
  Eigenstate&$\tilde{u}_{1 L}$ & $\tilde{u}_{1 R}$ & $\tilde{u}_{2
   L}$
   &$ \tilde{u}_{2 R}$&$\tilde{u}'_{L}$
   &$\tilde{u}'_{R}$
   \\
\hline
  $(\textrm{ Mass})^2$ & $\overline{E}_{u_{1 L}}$ & $\overline{A}_{11}$
  &$ \overline{F}_{u_{22}}$ & $\overline{A}_{22}$
  & $\overline{E}_{u^\prime_L}$
  &$ \overline{C}$\\
   \hline
 \end{tabular}
 \label{eigensquarks}
\end{center}
\end{table}
and two  others are \bea \tilde{u}_{t_L} &=&
s_{\theta_u}\tilde{u}_{3 R}
- c_{\theta_u}\tilde{u}_{3 L}, \label{sf270a}\\
\tilde{u}_{t_R} &=& c_{\theta_u}\tilde{u}_{3
R}+s_{\theta_u}\tilde{u}_{3 L}, \label{sf270b}\eea with respective
masses \bea m^2_{\tilde{u}_{t L}}&=&  \fr 1 2 (\overline{F}_{u_{33}}
+ \overline{A}_{33} - \overline{\De}),\\
m^2_{\tilde{u}_{t R}}&=&  \fr 1 2 (\overline{F}_{u_{33}} +
\overline{A}_{33} + \overline{\De}), \label{sf270c}\eea where \bea
\overline{\De} &=& \sqrt{(\overline{A}_{33}-
\overline{F}_{u_{33}})^2
+ 4\overline{G}_{33}^2},\label{sf350}\\
t_{2\theta_u}&=&\fr{2\overline{G}_{33}}{\overline{A}_{33}
-\overline{F}_{u_{33}}}\label{sf360}\eea

\item {\it For down-squarks}:

Eigenstates and masses are presented in Table \ref{eigedownq}.
\begin{table}[h]
\caption{Masses and eigenstates of  down-squarks}
\begin{center}
 \begin{tabular}{|c|c|c|c|c|c|c|}
   \hline
  Eigenstate&$\tilde{d}_{1 L}$ & $\tilde{d}_{1 R}$ & $\tilde{d}_{2
   L}$
   &$ \tilde{d}_{2 R}$&$\tilde{d}'_{2L}$
   &$\tilde{d}'_{2R}$
   \\
\hline
  $(\textrm{ Mass})^2$ & $\overline{E}_{d_{1 L}}$
  & $\overline{B}_{11}$
  &$ \overline{F}_{d_{22}}$ & $\overline{B}_{22}$
  &$ \overline{F}_{d^\prime_{22}}$
  &$ \overline{P}_{22}$
  \\
   \hline
 \end{tabular}
 \label{eigedownq}
\end{center}
\end{table}
\een
and four  others are \bea \tilde{d}_{t_L} &=&
s_{\theta_d}\tilde{d}_{3 R}
- c_{\theta_d}\tilde{d}_{3 L}, \label{sf270a}\\
\tilde{d}_{t_R} &=& c_{\theta_d}\tilde{d}_{3
R}+s_{\theta_d}\tilde{d}_{3 L}, \label{sf270b}\eea with respective
masses \bea m^2_{\tilde{d}_{t L}}&=&  \fr 1 2 (\overline{F}_{d_{33}}
+ \overline{B}_{33} - \overline{\De}),\\
m^2_{\tilde{d}_{t R}}&=&  \fr 1 2 (\overline{F}_{d_{33}} +
\overline{B}_{33} + \overline{\De}), \label{sf270c}\eea where \bea
\overline{\De} &=& \sqrt{(\overline{B}_{33}-
\overline{F}_{d_{33}})^2
+ 4\overline{K}_{33}^2},\label{sf350}\\
t_{2\theta_d}&=&\fr{2\overline{K}_{33}}{\overline{B}_{33}
-\overline{F}_{d_{33}}}.\label{sf360}\eea
 Analogously for exotic
squarks:
 \bea \tilde{d}'_{t_L} &=&
s_{\theta_{d'}}\tilde{d}'_{3 R}
- c_{\theta_{d'}}\tilde{d}'_{3 L}, \label{sf270a}\\
\tilde{d}'_{t_R} &=& c_{\theta_{d'}}\tilde{d}'_{3
R}+s_{\theta_{d'}}\tilde{d}'_{3 L}, \label{sf270b}\eea with
respective masses \bea m^2_{\tilde{d}'_{t L}}&=&  \fr 1 2
(\overline{F}_{d'_{33}}
+ \overline{P}_{33} - \overline{\De}),\\
m^2_{\tilde{d}'_{t R}}&=&  \fr 1 2 (\overline{F}_{d'_{33}} +
\overline{P}_{33} + \overline{\De}), \label{sf270c}\eea where \bea
\overline{\De} &=& \sqrt{(\overline{P}_{33}-
\overline{F}_{d'_{33}})^2
+ 4\overline{N}_{33}^2},\label{sf350}\\
t_{2\theta_{d'}}&=&\fr{2\overline{N}_{33}}{\overline{P}_{33}
-\overline{F}_{d'_{33}}}.\label{sf360}\eea

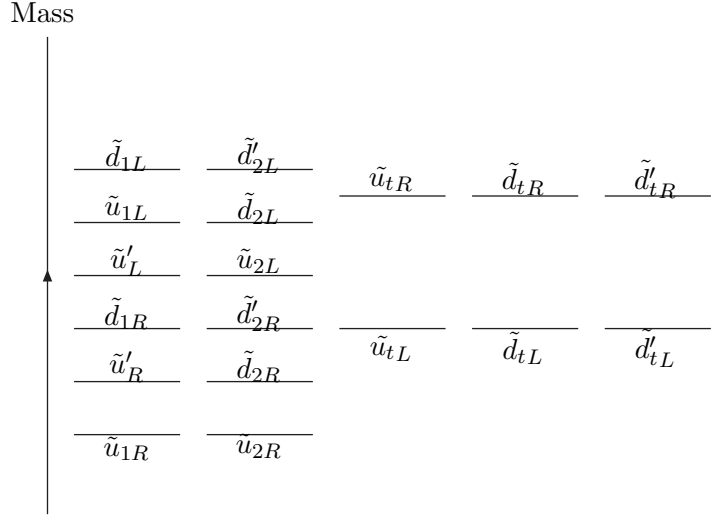
\begin{figure}[h]
\begin{center}
\begin{picture}(250,150)(0,0)
\ArrowLine(0,0)(0,180)
 \Text(0,190)[]{Mass }
 \Line(10,30)(50,30)
\Text(30,25)[]{$\tilde{u}_{1 R}$}
 \Line(10,50)(50,50)
\Text(30,56)[]{$\tilde{u}^\prime_{ R}$}
 \Line(10,70)(50,70)
\Text(30,76)[]{$\tilde{d}_{1 R}$}
 \Line(10,90)(50,90)
\Text(30,96)[]{$\tilde{u}^\prime_{L}$} \Line(10,110)(50,110)
\Text(30,116)[]{$\tilde{u}_{1 L}$} \Line(10,130)(50,130)
\Text(30,136)[]{$\tilde{d}_{1 L}$}
 \Line(60,30)(100,30)
\Text(80,26)[]{$\tilde{u}_{2 R}$}
 \Line(60,50)(100,50)
\Text(80,56)[]{$\tilde{d}_{2 R}$}
 \Line(60,70)(100,70)
\Text(80,76)[]{$\tilde{d}^\prime_{2 R}$}
 \Line(60,90)(100,90)
\Text(80,96)[]{$\tilde{u}_{2 L}$} \Line(60,110)(100,110)
\Text(80,116)[]{$\tilde{d}_{2 L}$} \Line(60,130)(100,130)
\Text(80,136)[]{$\tilde{d}^\prime_{2 L}$}
\Line(110,70)(150,70)\Line(110,120)(150,120)
\Text(130,63)[]{$\tilde{u_t}_{ L}$} \Text(130,127)[]{$\tilde{u_t}_{
R}$}
\Line(160,70)(200,70)\Line(160,120)(200,120)
\Text(180,63)[]{$\tilde{d_t}_{ L}$} \Text(180,127)[]{$\tilde{d_t}_{
R}$}
\Line(210,70)(250,70)\Line(210,120)(250,120)
\Text(230,63)[]{$\tilde{d'_t}_{ L}$}
\Text(230,127)[]{$\tilde{d'_t}_{ R}$}
\end{picture}
\end{center}
\caption[]{ A schematic sample mass spectrum for squarks in which
mass scales between generations are not taken into account. }
\label{Squarkspec2}
\end{figure}

 To outline mass spectrum, let us  assume
\bea && \overline{A}_{11} <\overline{C} <
\overline{B}_{11}<\overline{E}_{u^\prime_L} <\overline{E}_{u_{1 L}}
<\overline{E}_{d_{1 L}},\crn && \overline{A}_{22}<\overline{B}_{22}<
\overline{P}_{22}<\overline{F}_{u_{22}} < \overline{F}_{d_{22}} <
\overline{F}_{d^\prime_{22}}.
 \label{sf362}\eea
 Squarks  mass spectrum is
shown in figure \ref{Squarkspec2}, where mass scales between
generations are not taken into account. \vspace{2cm}

We summarize  this section by notice that the  huge squark mixing
matrices ($8 \times 8$ and $10 \times 10$) were significally reduced
by the lepton number conservation. The situation will be much better
by $R$-parity imposition.

\section{\label{rparity}$R$-parity and sfermion mass splitting  }

Consequence of $R$-parity is that all coefficients in
$W_{R\!\!\!\!/}$ and $\mathcal{L}^{R\!\!\!\!/}_{SMT}$ vanish.

\subsection{Slepton  mass splitting}

$R$-parity conservation and the constraint (\ref{hsyln}) give \bea
\la_a=\la'_{ab}=M'^2_a=\upsilon_a=\mu_{0a}=0.\eea Then vanishing of
nondiagonal
 elements in lepton mixing  matrices leads to:
\bea \ga_{c1}\ga_{c2}=\ga_{c1}\ga_{c3}=\ga_{c3}\ga_{c2}=0.
\label{th1}\eea  Consequence of (\ref{th1}) is that at least, one
of the coefficients $\ga_{ab}$ vanishes. Let us consider two
 special cases: \ben
\item  $\ga_{c 3} \neq 0$

From (\ref{th1}) we get \be \ga_{c1}= \ga_{c2} = 0, \hs \Rightarrow
\hs \ga_{33} \neq 0. \label{sf49} \ee In the considering  case, the
first two family slepton  masses are given by: \bea
 m^2_{\tilde{l}_{1 L}} & = & M^2_{11}
 - \fr{g^2}{2} \left( H_3
 - \fr{1}{\sqrt{3}}H_8 + \fr{2 t^2}{3}H_1 \right),\label{sf66}\\
 m^2_{\tilde{\nu}_{1 L}} & = &  M^2_{11} +
 \fr{g^2}{2} \left( H_3
 + \fr{1}{\sqrt{3}}H_8 - \fr{2 t^2}{3}H_1 \right),\label{sf661}\\
m^2_{\tilde{\nu}_{1 R}}& = & M^2_{11}  -g^2 \left(
\fr{1}{\sqrt{3}}H_8 +
 \fr{ t^2}{3}H_1 \right),\label{sf663}\\
  m^2_{\tilde{l}_{1R}} & = & m^2_{11} +
   g^2 t^2 H_1,\label{sf662}\\
 m^2_{\tilde{l}_{2 L}} & = &  M^2_{22}
  - \fr{g^2}{2} \left( H_3  - \fr{1}{\sqrt{3}}H_8 + \fr{2
t^2}{3}H_1
\right),\label{sf664}\\
 m^2_{\tilde{\nu}_{2 L}}  & = &  M^2_{22} + \fr{g^2}{2} \left( H_3
 + \fr{1}{\sqrt{3}}H_8 - \fr{2 t^2}{3}H_1 \right),\label{sf665}\\
 m^2_{\tilde{\nu}_{2 R}}  & = &  M^2_{22} - g^2 \left( \fr{1}{\sqrt{3}}H_8 +
 \fr{ t^2}{3}H_1 \right),\label{sf666}\\
 m^2_{\tilde{l}_{2R}} & = & m^2_{22} +  g^2 t^2 H_1,\label{sf667}
 \eea The stau masses  are defined:
 \bea
m^2_{\tilde{\tau}_{ L}} & = & \fr {1}
{2}\left[M_{33}^2+m_{33}^2+\fr{v^{\prime 2}}{9} \ga_{33}^2 -
\fr{g^2}{2} \left( H_3
 - \fr{1}{\sqrt{3}}H_8 - \fr{2 t^2}{3}H_1 \right)
 -\De_1\right],\label{sf671}\\
 m^2_{\tilde{\tau}_{ R}} & = & \fr {1}
{2}\left[M_{33}^2+m_{33}^2+\fr{v^{\prime 2}}{9} \ga_{33}^2 -
\fr{g^2}{2} \left( H_3
 - \fr{1}{\sqrt{3}}H_8 - \fr{2 t^2}{3}H_1 \right)+\De_1\right],
 \label{sf68} \eea where \be \De_1 =
\sqrt{\left[M_{33}^2 -m_{33}^2  -  g^2 \left( H_3
 - \fr{1}{\sqrt{3}}H_8 + \fr{8 t^2}{3}H_1 \right)\right]^2
 + 2 \left(\eta_{33} v^\prime + \fr 1
6 \mu_\rho \ga_{33}v\right)^2}.\label{sf69}\ee

 For the mixing
sneutrino eigenstates \bea \tilde{\nu}_{\tau L} &=&
s_{\theta_n}\tilde{\nu}_{3 R}
- c_{\theta_n}\tilde{\nu}_{3 L}, \label{sf27an}\\
\tilde{\nu}_{ \tau R} &=& c_{\theta_n}\tilde{\nu}_{3
R}+s_{\theta_n}\tilde{\nu}_{3 L}, \label{sf27bn}\eea
 we obtain \bea  m^2_{\tilde{\nu}_{\tau L}} & =
& \fr {1} {2}\left[2M_{33}^2+ \fr{g^2}{2} \left( H_3
 - \fr{1}{\sqrt{3}}H_8 - \fr{4 t^2}{3}H_1 \right)
-\De_{n1}\right],\label{sf44n1}\\
 m^2_{\tilde{\nu}_{\tau R}} & = & \fr {1}
{2}\left[2M_{33}^2+ \fr{g^2}{2} \left( H_3
 - \fr{1}{\sqrt{3}}H_8 - \fr{4 t^2}{3}H_1 \right)
+\De_{n1}\right],\label{sf45n1} \eea with \be \De_{n1} = \sqrt{
\fr{g^4}{4} \left( H_3
 + \sqrt{3}H_8  \right)^2
+ 8 \left(\varepsilon_{33} v + \fr 1 6 \mu_\rho
\la^\prime_{33}v^\prime\right)^2}.\label{sf6n1}\ee

From Eq. (\ref{sf66}) to Eq. (\ref{sf667}), the mass splittings for
the sleptons are governed by sum-rules \bea m^2_{\tilde{l}_{1 L}} -
m^2_{\tilde{\nu}_{1 L}}  & = & m^2_{\tilde{l}_{2 L}} -
m^2_{\tilde{\nu}_{2 L}} = -  g^2 H_3 = \fr{g^2}{4} \left( v^2
\fr{\cos 2\ga}{c_\ga^2} + u^2 \fr{\cos 2\bet}{s_\bet^2} \right)\crn
& = & m_W^2 \cos 2 \ga + \fr{g^2 u^2}{4}\fr{\cos 2\bet}{s_\bet^2},
\label{sf201}\\
m^2_{\tilde{\nu}_{1 L}} - m^2_{\tilde{\nu}_{1 R}} & = &
m^2_{\tilde{\nu}_{2 L}} - m^2_{\tilde{\nu}_{2 R}} = \fr{g^2}{2}
\left( H_3 +\sqrt{3} H_8\right) =  \fr{g^2}{ 4} (w^2 -u^2) \fr{\cos
2\bet}{s_\bet^2}  \label{sf202}
 \eea

Remind that, in the effective approximation, we have
\cite{susyeco,{higph}}: $w \simeq w^\prime, u \simeq u^\prime$.
Thus,  noting that our notation $\tan\ga$ is $\cot\bet$ in MSSM as
in Ref. \cite{martin}, Eq. (\ref{sf201}) coincides with the MSSM
result \cite{martin}. In this approximation, there is degeneration
among $\tilde{\nu}_{1(2) L}$ and $\tilde{\nu}_{1(2) R}$. As in the
MSSM, $\cos 2 \ga > 0$ in the allowed range $\tan \ga < 1$, we get
then $m^2_{\tilde{l}_{ L}} > m^2_{\tilde{\nu}_{l L}},\ l = e, \mu$.
Assuming further $\cos 2 \bet >0$, we obtain: $m^2_{\tilde{\nu}_{l
L}}
> m^2_{\tilde{\nu}_{l R}}$.

\begin{figure}[h]
\begin{center}
\begin{picture}(250,200)(0,0)
\ArrowLine(0,0)(0,180)
 \Text(0,190)[]{Mass }
\Line(10,50)(30,50) \Line(10,55)(30,55) \Line(10,48)(30,48)
\Text(25,65)[]{$\tilde{l}_{ L}$ } \Text(42,50)[]{$ \tilde{\nu}_{l
L}$ } \Text(25,38)[]{$ \tilde{\nu}_{l R}$ }
\Line(75,50)(95,50)
\Line(75,60)(95,60) \Text(85,40)[]{$\tilde{\nu}_{\tau L}$ }
\Text(85,70)[]{$\tilde{\nu}_{\tau R}$ }
\Line(135,50)(160,50) \Line(135,60)(160,60)
 \Text(150,70)[]{$\tilde{\mu}_{R} $}
\Text(150,40)[]{$\tilde{e}_{R} $}
\Line(185,55)(210,55) \Line(185,90)(210,90)
 \Text(200,100)[]{$\tilde{\tau}_{R} $}
\Text(200,45)[]{$\tilde{\tau}_{L} $}
\end{picture}
\caption[]{ A schematic sample mass spectrum for sleptons, in which
mass scales between generations are not taken into account and  $l =
e, \mu $. }
 \label{slepspec1}
\end{center}
\end{figure}
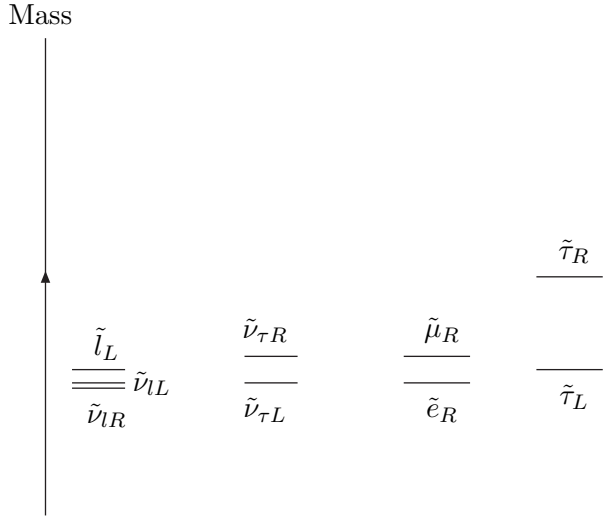

 To outline slepton mass spectrum, we
assume the following relationship: \bea && \cos 2\ga > 0,\hs \cos 2
\bet >0, \hs m^2_{11} < m^2_{22}.
 \eea With the above assumption, the slepton  mass
spectrum   is  shown in figure \ref{slepspec1}. Since, no convincing
evidence for production of superpartners has been found, our figure
has only illustrative meaning. \vs

\vs

\item  $\ga_{c 1} \neq 0$

As before, from (\ref{th1}) we have \be \ga_{c2}= \ga_{c3} = 0, \hs
\Rightarrow \hs \ga_{11} \neq 0. \label{sf74} \ee In this case,  all
the charged sleptons have different masses: \bea
 m^2_{\tilde{l}_{1 L}} & = & M^2_{11} +
 \fr{v^{\prime 2}}{18}\ga_{11}^2 - \fr{g^2}{2} \left( H_3
 - \fr{1}{\sqrt{3}}H_8 + \fr{2 t^2}{3}H_1 \right),\label{sf75}\\
 m^2_{\tilde{l}_{2 L}} & = &  M^2_{22} - \fr{g^2}{2} \left( H_3
 - \fr{1}{\sqrt{3}}H_8 + \fr{2 t^2}{3}H_1 \right),\label{sf76}\\
 m^2_{\tilde{l}_{1 R}} & = & m^2_{11} +
 \fr{v^{\prime 2}}{18}\ga_{11}^2 +  g^2 t^2 H_1 ,\label{sf77}\\
 m^2_{\tilde{l}_{2 R}} & = & m^2_{22} +  g^2 t^2 H_1 .
\label{sf78}\eea and \bea  m^2_{\tilde{\tau}_{ L}} & = & \fr {1}
{2}\left[M_{33}^2+m_{33}^2- \fr{g^2}{2} \left( H_3
 - \fr{1}{\sqrt{3}}H_8 - \fr{2 t^2}{3}H_1 \right)
-\De_2\right],\label{sf79}\\
 m^2_{\tilde{\tau}_{ R}} & = & \fr {1}
{2}\left[M_{33}^2+m_{33}^2 - \fr{g^2}{2} \left( H_3
 - \fr{1}{\sqrt{3}}H_8 - \fr{2 t^2}{3}H_1
 \right)  +\De_2
\right]\label{sf80} \eea with \be \De_2 = \sqrt{\left[M_{33}^2
-m_{33}^2 - \fr{g^2}{2} \left( H_3
 - \fr{1}{\sqrt{3}}H_8 + \fr{8 t^2}{3}H_1
 \right)\right]^2 + 2 \eta^2_{33} v^{\prime
2}}.\label{sf81}\ee

 For sneutrinos, we have
  \bea
 m^2_{\tilde{\nu}_{1 L}} & = &  M^2_{11} + \fr{g^2}{2} \left( H_3
 + \fr{1}{\sqrt{3}}H_8 - \fr{2 t^2}{3}H_1 \right),\label{sf82}\\
 m^2_{\tilde{\nu}_{1 R}} & = &  M^2_{11} -g^2
\left( \fr{1}{\sqrt{3}}H_8 +
 \fr{ t^2}{3}H_1 \right),\label{sf82}\\
 m^2_{\tilde{\nu}_{2 L}}
 & = & M^2_{22} + \fr{g^2}{2} \left( H_3
 + \fr{1}{\sqrt{3}}H_8 - \fr{2 t^2}{3}H_1 \right),\label{sf821}\\
 m^2_{\tilde{\nu}_{2 R}} & = & M^2_{22}  - g^2 \left( \fr{1}{\sqrt{3}}H_8 +
 \fr{ t^2}{3}H_1 \right).\label{sf83}
 \eea

For the tau sneutrinos, we get
 \bea  m^2_{\tilde{\nu}_{\tau L}} & = & \fr {1}
{2}\left[2 M_{33}^2+ \fr{g^2}{2} \left( H_3
 - \fr{1}{\sqrt{3}}H_8 - \fr{4 t^2}{3}H_1 \right)
-\De_{n}\right],\label{sf44n1}\\
 m^2_{\tilde{\nu}_{\tau R}} & = & \fr {1}
{2}\left[2M_{33}^2 + \fr{g^2}{2} \left( H_3
 - \fr{1}{\sqrt{3}}H_8 - \fr{4 t^2}{3}H_1 \right)
+\De_n\right],\label{sf45n1} \eea with \be \De_n = \sqrt{
\fr{g^4}{4} \left( H_3
 + \sqrt{3}H_8  \right)^2
+ 8 \left(\varepsilon_{33} v + \fr 1 6 \mu_\rho
\la^\prime_{33}v^\prime\right)^2}.\label{sf46n1}\ee In the
considered case, only relation among sneutrinos (\ref{sf202}) is
satisfied.

 In this case, slepton  mass spectrum
is similar to  figure \ref{slepspec1}. \een The situation is similar
for case of $\ga_{c 2} \neq 0$, in which the second generation plays
a role of the first one.
 Next, let us consider squarks in the model under
consideration.

\subsection{Squark  mass mixing matrices}

Imposition of $R-parity$ yields \bea \xi_{a\al i}=\xi'_{a\al
\bet}=\la_a=0. \eea Looking at (\ref{th2})--(\ref{th3}) we have
\bea \overline{K}_{\al i}= \overline{N}_{\al \bet} =
0\label{ng129}.\eea So that the mass matrix of down-squarks is
significally reduced.

With the help of (\ref{ng129}), matrix in (\ref{sf93s}) is
decomposed to two $2 \times 2$ ones. $\tilde{d}'^c_{2L}$ and
$\tilde{d}'^c_{3L}$ mix with mass matrix \bea \left(
   \begin{array}{cc}
     \overline{P}_{22} & \overline{P}_{23} \\
     \overline{P}_{32} & \overline{P}_{33}
   \end{array}
 \right).
\label{sf93} \eea For $\tilde{d}'_{2L}$ and $\tilde{d}'_{3L}$, the
mass matrix is \bea \left(
   \begin{array}{cc}
     \overline{F}_{d^\prime_{2 2}} & \overline{F}_{d^\prime_{2 3}} \\
     \overline{F}_{d^\prime_{3 2}} & \overline{F}_{d^\prime_{33}}
   \end{array}
 \right).
 \label{sf94}
 \eea
 For the ordinary down-squark, matrix  in (\ref{sf98s}) is decomposed
 into $2 \times 2$ and $4 \times 4$ ones.  We have two blocks: in the base
 $(\tilde{d}_{2L},\tilde{d}_{3L})$, the mass matrix is given by
\bea \left(
  \begin{array}{cc}
    \overline{F}_{d_{22}} & \overline{F}_{d_{23}} \\
    \overline{F}_{d_{32}} & \overline{F}_{d_{33}} \\
  \end{array}
\right).\label{sf97} \eea Four others mix and, in the base
$(\tilde{d}^*_{1L},\tilde{d}^c_{1L},
 \tilde{d}^c_{2L},\tilde{d}^c_{3L})$, the mass matrix is defined
by \bea \left(
  \begin{array}{cccc}
     \overline{E}_{d_{1L}}& \overline{H}_1 &
     \overline{H}_2  & \overline{H}_3 \\
     \overline{H}_1 & \overline{B}_{11}
      & \overline{B}_{21} & \overline{B}_{31} \\
    \overline{H}_2 & \overline{B}_{12} &
    \overline{B}_{22} & \overline{B}_{32} \\
    \overline{H}_3 & \overline{B}_{13}
    & \overline{B}_{23} & \overline{B}_{33} \\
  \end{array}
\right).\label{sf98} \eea It is interesting to note that our
highest mass mixing matrices are smaller than that in the MSSM
($6 \times 6$ matrices).

Let us consider the squark  mass splitting.  Looking at
Eqs.(\ref{cte})-(\ref{ctft}) yields the mass splitting of
 squarks in the first  generation:
\bea m^2_{\tilde{d}_{1 L}} - m^2_{\tilde{u}_{1 L}}& = &
\overline{E}_{d_{1 L}} - \overline{E}_{u_{1 L}} = - g^2 H_3
+\fr{1}{18}\left[v'^2(\vartheta^2_i +\vartheta'^2_\al)-u'^2
(\kappa^2_i +\kappa'^2)\right]\crn &=& \fr{g^2}{4} \left( v^2
\fr{\cos 2\ga}{c_\ga^2} + u^2 \fr{\cos 2\bet}{s_\bet^2} \right)
+\fr{1}{18}\left[v'^2(\vartheta^2_i +\vartheta'^2_\al)-u'^2
(\kappa^2_i +\kappa'^2)\right]\crn & =& m_W^2 \cos 2\ga + \fr{g^2
u^2}{4} \fr{\cos 2\bet}{s_\bet^2}+\fr{1}{18}\left[v'^2(\vartheta^2_i
+\vartheta'^2_\al)-u'^2 (\kappa^2_i +\kappa'^2)\right]\crn
  &\simeq &  m_W^2 \cos
2\ga +\fr{1}{18}v'^2 \vartheta^2_i.\label{sf201q}\\
  \eea
Similarly, for the second generation, we get \bea
m^2_{\tilde{d}_{2 L}} - m^2_{\tilde{u}_{2 L}}& =
&\overline{F}_{d_{2 2}} - \overline{F}_{u_{22}} = - g^2 H_3
\crn&&+ \fr {1}{18}[u^2(\Pi'_{\bet\de}\Pi'_{\al\de}+\Pi_{\bet
i}\Pi_{\al i })-v^2(\pi_{\bet i}\pi_{\al i }+\pi'_\bet
\pi'_\al)]\crn
 &\simeq &  m_W^2 \cos  2\ga -
 \fr {1}{18}v^2 \pi_{\bet i}\pi_{\al i }.
 \label{hqtkl} \eea

In the model under consideration, squark mass splitting is
different from those in the MSSM and the reason of this is the
quark generation discrimination. For sleptons, the splitting is
the same as in the MSSM. In addition, in the SM limit, we have
triple degeneracy among all particles in the lepton triplet.

\section{\label{concl} Conclusions}

In this paper we have studied the sfermion sector in the
supersymmetric economical 3-3-1 model. Our calculation of the full
superpotential for sfermions is useful for further study on
searching of supersymmetric particles at high energy colliders such
as the CERN Large Hadron Collider (LHC),... By $R$- parity
conservation imposition, the Higgs scalars are decoupled of the
sfermions; and the exotic squarks are also decoupled of
superpartners of the ordinary quarks.

In contradiction to the MSSM, in the model under consideration,
there are lepton number violating mass terms in the  contribution
from $D$-part.

As in the MSSM, the mass mixing matrix for charged sleptons is $6
\times 6$, while for sneutrinos, due to the existence of the
right-handed neutrinos, their mass mixing matrix is $6 \times 6$ too
(Remind that in the MSSM, it is $3 \times 3$ matrix).

It is worth noting that, in the SM limit,  due to the quark family
discrimination, the highest mass mixing matrices for the up-squarks
and the down-squarks are, respectively,  $5 \times 5$ and $4 \times
4$, but not $6 \times 6$ as in the MSSM. Due to the same reason, in
contradiction with the MSSM, there is no mixing among $\tilde{b}_L$
and $\tilde{b}_R$.

Assuming that there is only mixing among highest flavors (
 $\tilde{\nu}_{\tau L} - \tilde{\nu}_{\tau R}$,
 $\tilde{\tau}_{ L} - \tilde{\tau}_{ R}$
and $\tilde{t}_{ L} - \tilde{t}_{ R}$)
    we were able to outline
 mass spectra for the sfermions in the model.

In the SM limit, without $D$-term contribution, there is triple
degeneracy among all particles in the lepton triplet. Therefore
the mass splitting among sleptons is proportional the $D$-term
contribution \bea m^2_{\tilde{l}_{ L}} - m^2_{\tilde{\nu}_{ L}} &
\simeq &  m^2_W \cos 2 \ga ,\crn  m^2_{\tilde{\nu}_{e L}} & = &
m^2_{\tilde{\nu}_{e R}}.\label{hq2} \eea

However, due to the quark generation discrimination, squark mass
splittings are different in each family and from those in the
MSSM.

 We do hope that, in coming years, the CERN LHC will provide
important information on the supersymmetric particles including
sfermions and our prediction in Eq. (\ref{hq2}) can be
experimentally checked.

To conclude this work, we note again that due to the minimal content
of the scalar sector, in the supersymmetric  economical 3-3-1 model,
Higgs sector is quite constrained and the significant number of free
parameters is reduced. Its supersymmetric extension has the same
feature and deserves further studies.

\section*{Acknowledgments}
The work was supported in part by National Council for Natural
Sciences of Vietnam under grant  No: 410604.
\\[0.3cm]

\appendix

\section{The F-term contribution}

Here we present full $F$-term contributions to sfermion masses:
\bea
 F^{\chi^\prime *}F_{\chi^\prime} & = & \fr 1 4 \mu_{0a} \mu_{0b} \tilde{L}_{aL}^*
\tilde{L}_{bL} + \fr 1 4 \mu_{0a} \mu_{\chi} (\tilde{L}_{aL}^* \chi
+ \tilde{L}_{aL} \chi^*)\crn && + \fr 1 6 \mu_\chi \chi^*
(\kappa_{i} \tilde{Q}_{1L} \tilde{u}^{c}_{iL} + \kappa^\prime
\tilde{Q}_{1L} \tilde{u}^{\prime c}_L ) + \fr 1 6 \mu_\chi \chi
(\kappa_{i} \tilde{Q}_{1L}^* \tilde{u}^{c *}_{iL} + \kappa^\prime
\tilde{Q}_{1L}^* \tilde{u}^{\prime c *}_L ),\\
  F^{\chi_\si *} F_{\chi^\si} & = & \left[\fr 1 6 \mu_\chi
\chi^{\prime \si *} \left(\la_{a} \epsilon_{m \si n} \tilde{L}_{a
L}^m \rho^n + \Pi_{\alpha i} \tilde{Q}_{\alpha L \si}
\tilde{d}^{c}_{iL} + \Pi^\prime_{\alpha \bet} \tilde{Q}_{\alpha L
\si} \tilde{d}^{\prime c}_{\bet L}\right)+H.c.\right]\crn && + \fr1
9 \la_a \la_b [(\tilde{L}^*_{aL}\tilde{L}_{bL})(\rho^*\rho)
-(\tilde{L}^*_{aL}\rho)(\rho^*\tilde{L}_{bL})]\crn & = & \left[\fr 1
6 \mu_\chi \left(\la_{a} \epsilon \tilde{L}_{a L}\chi^{\prime *}
\rho + \Pi_{\alpha i} \chi^{\prime *}. \tilde{Q}_{\alpha L }
\tilde{d}^{c}_{iL} + \Pi^\prime_{\alpha \bet} \chi^{\prime *}.
\tilde{Q}_{\alpha L } \tilde{d}^{\prime c}_{\bet L}\right) +
H.c.\right]\crn && + \fr1 9 \la_a \la_b
[(\tilde{L}^*_{aL}\tilde{L}_{bL})(\rho^*\rho)
-(\tilde{L}^*_{aL}\rho)(\rho^*\tilde{L}_{bL})],\\
F^{\rho_\si *} F_{\rho^\si} & = & \left[\fr 1 6 \mu_\rho
\rho^{\prime \si *} \left(\la_{a} \epsilon_{m n \si} \tilde{L}_{a
L}^m \chi^n + \la^\prime_{ab} \epsilon_{m n \si } \tilde{L}_{a L}^m
\tilde{L}_{b L}^n  \right.\right. \crn && + \left. \left. \pi_{
\alpha i} \tilde{Q}_{\alpha L \si}\tilde{u}^{c}_{iL}
+\pi_{\alpha}^{\prime} \tilde{Q}_{\alpha L \si}\tilde{u}^{\prime
c}_{L}\right) + H.c.\right]\crn && + \fr1 9 \la_a \la_b
[(\tilde{L}^*_{aL}\tilde{L}_{bL})(\chi^*\chi)
-(\tilde{L}^*_{aL}\chi)(\chi^*\tilde{L}_{bL})] \crn & = &\left[ \fr
1 6 \mu_\rho \left( \la_{a} \epsilon \tilde{L}_{a L} \chi
\rho^{\prime *}+ \la^\prime_{ab} \epsilon \tilde{L}_{a L}
\tilde{L}_{b L} \rho^{\prime *}
 \right.\right. \crn && + \left. \left. \pi_{ \alpha i}
 \rho^{\prime *}. \tilde{Q}_{\alpha L }\tilde{u}^{c}_{iL} +
\pi_{\alpha}^{\prime}  \rho^{\prime *}.\tilde{Q}_{\alpha L
}\tilde{u}^{\prime c}_{L}
 \right) + H.c.\right]\crn && + \fr1 9 \la_a \la_b
[(\tilde{L}^*_{aL}\tilde{L}_{bL})(\chi^*\chi)
-(\tilde{L}^*_{aL}\chi)(\chi^*\tilde{L}_{bL})],\\
 F^{\rho^\prime *} F_{\rho^\prime} & = & \fr 1 6 \mu_\rho
\rho^* \left(\ga_{ab} \tilde{L}_{a L} \tilde{l}_{b L}^c +
\vartheta_{i}\tilde{Q}_{1L} \tilde{d}^{c}_{iL} + \vartheta^\prime_{
\alpha}\tilde{Q}_{1L} \tilde{d}^{\prime
c}_{\alpha L} \right) + H.c. \label{sf10},\\
F^{L^{\si}_{a L}*}F_{L^\si_{a L}} & = & \left[\fr 1 6  \mu_{0a}
 \chi^{\prime \si *} \left(\ga_{ab}  \rho^{\prime}_\si \tilde{l}_{b L}^c +
\la_{a} \epsilon_{\si m n}  \chi^m \rho^n+ 2 \la^\prime_{ab}
\epsilon_{\si m n}  \tilde{L}^m_{bL} \rho^n \right. \right.\crn && +
\left.   \xi_{a \alpha j} \tilde{Q}_{\alpha L \si}
\tilde{d}^{c}_{jL}+ \xi^\prime_{a\alpha \beta} \tilde{Q}_{\alpha L
\si} \tilde{d}^{\prime c}_{\beta L}\right)   \crn && + \fr 1 9
\left(\ga_{ab}\la_a  \epsilon\rho^{\prime } \chi^*\rho^*.
\tilde{l}_{b L}^c + 2\ga_{ab} \la^\prime_{ab} \epsilon\rho'
\tilde{L}^*_{bL} \rho^*  .\tilde{l}_{b L}^c\right. \crn &&
+2\la_a\la'_{ab}
[(\chi^*\tilde{L}_{bL})(\rho^*\rho)-(\rho^*\tilde{L}_{bL})(\chi^*\rho)]
\crn &&\left.\left.+\la_a \xi_{a \alpha j} \epsilon\tilde{Q}_{\alpha
L }\chi^*\rho^* . \tilde{d}^{c}_{jL}+ \la_a\xi^\prime_{a\alpha
\beta} \epsilon\tilde{Q}_{\alpha L }\chi^*\rho^* . \tilde{d}^{\prime
c}_{\beta L}\right)+H.c.\right]\crn&&+\fr 1 9 \ga_{c a} \ga_{c b}
\rho'^*. \rho'  \tilde{l}_{a L}^{c*}\tilde{l}_{b L}^c\crn&&+\fr 4 9
\la^\prime_{ca}\la^\prime_{cb}[(\tilde{L}^*_{aL}\tilde{L}_{bL})(\rho^*\rho)
-(\tilde{L}^*_{aL}\rho)(\rho^*\tilde{L}_{bL})],\\
 F^{l^{c}_{L b }*} F_{l^{c}_{L b}}& = & \fr 1 9 \ga_{ab} \ga_{a'b}
(\tilde{L}_{aL} \rho^{\prime}) (\tilde{L}_{a'L} \rho^{\prime})^*,\\
 F^{Q_{1 L}*}  F_{Q_{1 L}} & = & \fr {1} {18} [(w^{\prime 2} + u^{\prime 2})
  (\kappa_{i}\kappa_{j}
\tilde{u}^{c *}_{iL}\tilde{u}^{c}_{jL} + \kappa^{\prime 2}
\tilde{u}^{\prime c *}_L \tilde{u}^{\prime c}_L +
\kappa_{i}\kappa^\prime\tilde{u}^{c *}_{iL}\tilde{u}^{\prime c}_L +
\kappa_{i}\kappa^\prime\tilde{u}^{c }_{iL}\tilde{u}^{\prime c
*}_L ) \crn && + v^{\prime 2}(\vartheta_{i}\vartheta_{j}
\tilde{d}^{c *}_{iL}\tilde{d}^{c}_{jL} + \vartheta_{ \al}^{\prime
}\vartheta_{ \bet}^{\prime }\tilde{d}^{\prime c *}_{\al L}
\tilde{d}^{\prime c}_{\bet L}  + \vartheta_{i} \vartheta_{
\al}^{\prime }\tilde{d}^{c *}_{iL}\tilde{d}^{\prime c}_{\al L} +
\vartheta_{i} \vartheta_{ \al}^{\prime }\tilde{d}^{c
}_{iL}\tilde{d}^{\prime c *}_{\al L}),\\
 F^{Q_{\al L}*}  F_{Q_{\al L}} & = & \fr {1}{18}[v^2(\pi_{ \al i}\pi_{ \al j}
\tilde{u}^{c *}_{iL} \tilde{u}^{c }_{jL} + \pi_{\al}^{\prime
2}\tilde{u}^{\prime c *}_L \tilde{u}^{\prime c}_L + \pi_{ \al
i}\pi_{ \al}^{\prime}\tilde{u}^{c *}_{iL} \tilde{u}^{\prime c}_L +
\pi_{ \al i}\pi_{ \al}^{\prime}\tilde{u}^{c }_{iL} \tilde{u}^{\prime
c *}_L
 )\crn && + (w^{ 2} + u^{ 2})(\Pi_{ \al i}\Pi_{ \al j}
 \tilde{d}^{ c *}_{iL}\tilde{d}^{ c}_{jL} + \Pi_{ \al \bet}^{\prime }
 \Pi_{ \al \de}^{\prime }\tilde{d}^{\prime c *}_{\bet L}
 \tilde{d}^{\prime c}_{\de L}\crn && + \Pi_{ \al i}\Pi_{  \al \bet}^{\prime }
 \tilde{d}^{ c *}_{iL}\tilde{d}^{\prime c}_{\bet L} +
\Pi_{ \al i}\Pi_{  \al \bet}^{\prime } \tilde{d}^{ c
}_{iL}\tilde{d}^{\prime c *}_{\bet L}  )] + \cdot \cdot \cdot,\\
F^{u_{i L}^{c}*}F_{u_{i L}^{c}} & = & \fr 1 9
\left[\kappa_{i}^2(\chi^\prime \widetilde{Q}_{1L})(\chi^\prime
\widetilde{Q}_{1L})^* +\pi_{ \al i} \pi_{ \bet i} (\rho
\widetilde{Q}_{\al L})(\rho \widetilde{Q}_{\bet L})^* \right.\crn
&&\left. + \kappa_{i}\pi_{ \al i}(\chi^\prime
\widetilde{Q}_{1L})(\rho \widetilde{Q}_{\al L})^* +
\kappa_{i}\pi_{ \al i}(\chi^\prime \widetilde{Q}_{1L})^*(\rho
\widetilde{Q}_{\al L})\right] + \cdot \cdot \cdot
\crn &=&\fr {1}{18} \left[u'^2\kappa_{i}^2
\widetilde{u}_{1L}^*\widetilde{u}_{1L}+  w'^2\kappa_{i}^2
\widetilde{u}'^*_{L}\widetilde{u}'_{L} +v^2 \pi_{ \al i} \pi_{
\bet i} \widetilde{u}_{\al L}\widetilde{u}^*_{\bet L}\right.\crn
&&\left.+(u' w' \kappa_{i}^2\widetilde{u}_{1L}^*\widetilde{u'}_{L}
-u' v \kappa_{i}\pi_{ \al i}\widetilde{u}_{1L}^*\widetilde{u}_{\al
L} -w'v \kappa_{i}\pi_{ \al
i}\widetilde{u}'^*_{L}\widetilde{u}_{\al
L} +H.c. )\right] + \cdot \cdot \cdot,\\
F^{u_{ L}^{'c}*}F_{u_{ L}^{'c}} & = &\fr {1}{18}
\left[u'^2\kappa'^2 \widetilde{u}_{1L}^*\widetilde{u}_{1L}+
w'^2\kappa'^2 \widetilde{u}'^*_{L}\widetilde{u}'_{L} +v^2 \pi'_{
\al } \pi'_{ \bet } \widetilde{u}_{\al L}\widetilde{u}^*_{\bet
L}\right.\crn &&\left.+(u' w'
\kappa'^2\widetilde{u}_{1L}^*\widetilde{u'}_{L} -u' v
\kappa'\pi'_{ \al}\widetilde{u}_{1L}^*\widetilde{u}_{\al L} -w'v
\kappa'\pi'_{ \al}\widetilde{u}'^*_{L}\widetilde{u}_{\al L} +H.c.
)\right] + \cdot \cdot \cdot,\\
F^{d_{i L}^{c}*}F_{d_{i L}^{c}} & = & \fr {1}{18} \left[v'^2
\vartheta_i^2 \widetilde{d}_{1L}\widetilde{d}_{1L}^* +u^2 \Pi_{\al
i}\Pi_{\bet i}\widetilde{d}_{\al L}\widetilde{d}^*_{\bet L} +w ^2
\Pi_{\al i}\Pi_{\bet i}\widetilde{d}'_{\al
L}\widetilde{d}'^*_{\bet L}+\right.\crn &&\left. +(u w \Pi_{\al
i}\Pi_{\bet i}\widetilde{d}_{\al L}\widetilde{d}'^*_{\bet L}+v'u
\vartheta_i\Pi_{\al i}\widetilde{d}_{1 L}\widetilde{d}^*_{\al
L}+v'w \vartheta_i\Pi_{\al i}\widetilde{d}_{1
L}\widetilde{d}'^*_{\al L} +H.c.)\right] + \cdot \cdot \cdot,\crn
 F^{d_{\de
L}^{'c}*}F_{d_{\de  L}^{'c}} & = & \fr {1}{18} \left[v'^2
\vartheta'^2_\de \widetilde{d}_{1L}\widetilde{d}_{1L}^* +u^2
\Pi'_{\al \de}\Pi'_{\bet \de }\widetilde{d}_{\al
L}\widetilde{d}^*_{\bet L} +w ^2 \Pi'_{\al \de}\Pi'_{\bet
\de}\widetilde{d}'_{\al L}\widetilde{d}'^*_{\bet L}+\right.\crn
&&\left. +(u w \Pi'_{\al \de}\Pi'_{\bet \de}\widetilde{d}_{\al
L}\widetilde{d}'^*_{\bet L}+v'u \vartheta'_\de\Pi'_{\al
\de}\widetilde{d}_{1 L}\widetilde{d}^*_{\al L}+v'w
\vartheta'_\de\Pi'_{\al \de}\widetilde{d}_{1
L}\widetilde{d}'^*_{\al L} +H.c.)\right] + \cdot \cdot \cdot.\nn
\eea
\end{document}